\documentclass[aoas]{imsart}

\RequirePackage{amsthm,amsmath,amsfonts,amssymb}
\RequirePackage[authoryear]{natbib}
\RequirePackage[colorlinks,citecolor=blue,urlcolor=blue]{hyperref}
\RequirePackage{graphicx}

\startlocaldefs
\theoremstyle{plain}
\newtheorem{theorem}{Theorem}
\newtheorem{lemma}{Lemma}
\usepackage{bm}
\usepackage{multirow}
\DeclareMathOperator*{\argmax}{argmax}

\endlocaldefs

\begin{document}

\begin{frontmatter}
\title{Semi-Confirmatory Factor Analysis for High-Dimensional Data with Interconnected Community Structures}
\runtitle{SCFA}

\begin{aug}
\author[A]{\fnms{Yifan}~\snm{Yang}\ead[label=e1]{yiorfun@case.edu} \orcid{0000-0001-5727-6540}},
\author[B]{\fnms{Tianzhou}~\snm{Ma}\ead[label=e2]{tma0929@umd.edu}}, 
\author[C]{\fnms{Chuan}~\snm{Bi}\ead[label=e3]{Chuan.Bi@som.umaryland.edu}}
\and
\author[C]{\fnms{Shuo}~\snm{Chen}\ead[label=e4]{shuochen@som.umaryland.edu}}
\address[A]{Department of Population and Quantitative Health Sciences, Case Western Reserve University \printead[presep={,\ }]{e1}}

\address[B]{Department of Epidemiology and Biostatistics, University of Maryland, College Park \printead[presep={,\ }]{e2}}

\address[C]{School of Medicine, University of Maryland \printead[presep={,\ }]{e3,e4}}
\end{aug}

\begin{abstract}
	Confirmatory factor analysis (CFA) is a statistical method for identifying and confirming the presence of latent factors among observed variables through the analysis of their covariance structure.  
	Compared to alternative factor models, CFA offers interpretable common factors with enhanced specificity and a more adaptable approach to covariance structure modeling. 
	However, the application of CFA has been limited by the requirement for prior knowledge about ``non-zero loadings'' and by the lack of computational scalability (e.g., it can be computationally intractable for hundreds of observed variables). 
	We propose a data-driven semi-confirmatory factor analysis (SCFA) model that attempts to alleviate these limitations.  
	SCFA automatically specifies ``non-zero loadings'' by learning the network structure of the large covariance matrix of observed variables, and then offers closed-form estimators for factor loadings, factor scores, covariances between common factors, and variances between errors using the likelihood method. 
	Therefore, SCFA is applicable to high-throughput datasets (e.g., hundreds of thousands of observed variables) without requiring prior knowledge about ``non-zero loadings''. 
	Through an extensive simulation analysis benchmarking against standard packages, SCFA exhibits superior performance in estimating model parameters with a much-reduced computational time.
	We illustrate its practical application through factor analysis on two high-dimensional RNA-seq gene expression datasets.
\end{abstract}


\end{frontmatter}


\section{Introduction}
	
\label{Sec:introduction}
	
	Factor analysis is a commonly used statistical technique to elucidate the relationship between multivariate observations. 
	Factor models aim to identify the underlying factors that collectively describe the interdependencies in multivariate observed data \citep{Anderson2003}.
	As correlated high-dimensional observed variables can be effectively decomposed into a smaller number of common factors, i.e., achieving dimension reduction, factor analysis models have garnered popularity in various fields, including social science, psychology, molecular biology, and others \citep{SchreiberNoraStage2006, FanLiZhang2020, FanWangZhong2021}.
	
	Factor analysis is broadly classified into two categories: exploratory factor analysis (EFA) and confirmatory factor analysis (CFA).
	EFA is frequently employed to explore the interactive relationships among observed variables and to identify latent common factors, without prerequisite knowledge of grouping these observed variables. 
	In contrast, CFA is commonly utilized to validate whether the empirical evidence supports a predetermined latent structure of the shared variance in the model specification. 
	In a CFA model, prerequisite knowledge or empirical evidence regarding the grouping of observed variables is represented by predefined ``non-zero loadings'' in the factor loading matrix, establishing a rule or a factor membership that exclusively and exhaustively assigns each observed variable to a certain common factor \citep{Browne2001}.  
	In practical applications, a combined approach is often adopted, starting with EFA to investigate the underlying dependence pattern, followed by CFA for model verification and justification \citep{Basilevsky2009, Brown2015, GanaBroc2019}.
	
	The above distinctive model specifications between EFA and CFA confer unique strengths and limitations to each approach. 
	EFA exhibits greater flexibility by not necessitating specified ``non-zero loadings'' and can be adapted to accommodate high-dimensional observations, reaching thousands of observed variables or more \citep{FriguetKloaregCauseur2009, BaiLi2012, FanLiaoMincheva2013, FanHan2017}. 
	However, the factor loadings in classical EFA models are typically non-zero, resulting in less interpretable relationships between common factors and observed variables.
	Moreover, EFA models are limited to explicitly estimating the covariance matrix of common factors for high-dimensional data.
	In contrast, CFA naturally specifies a sparse factor loading matrix guided by prior knowledge, enhancing interpretability, as exemplified in \citet{CarlsonMulaik1993}, and establishes arbitrary covariances between common factors, as demonstrated in  \citet{Lawley1958} and \citet{JacksonGillaspyJrPurcStephenson2009}.
	As previously mentioned, classical CFA models encounter two primary limitations: (1) predetermined ``non-zero loadings'' in the factor loading matrix or factor membership is typically lacking, resulting in the nonexistence of a rule that exclusively and exhaustively assigns each observed variable to a certain common factor; and (2) the computational burden of estimating a CFA model in high-dimensional scenarios becomes intractable because the existing standard computational packages struggle to handle datasets containing hundreds or thousands of observed variables \citep{Fox2006, Rosseel2012, Oberski2014}.
	
\begin{figure}[!htb]
		\centering
		\includegraphics[width = 1 \linewidth]{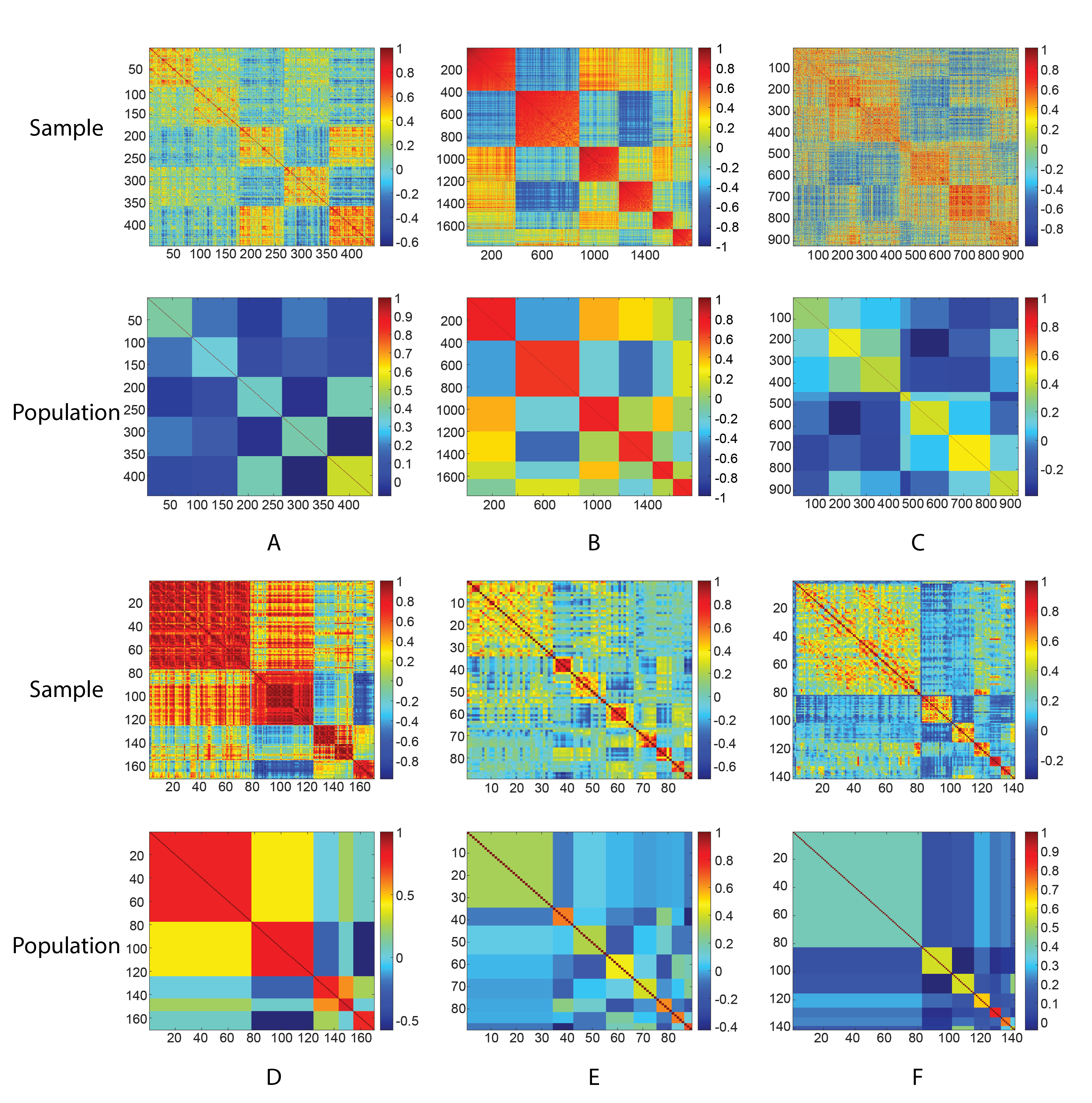} 
		\caption{
			Real-world examples demonstrated that the interconnected community structures are frequently observed:
			A: a brain imaging study \citep{ChiappelliRowlandWijtenburg2019};
			B: a gene expression study \citep{KatarzynaPatrycjaMaciej2015};
			C: a multi-omics study \citep{PerrotLevyRajjou2022}; 
			D: a plasma metabolomics study \citep{RitchieSurendranKarthikeyan2023};
			E and F: an environmental study involving exposome and plasma metabolomics \citep{ISG2021}.
		}
		\label{Fig:ics_examples}
	\end{figure}

	In the current research, we concentrate on the CFA approach while attempting to address the two aforementioned limitations.
	We propose a semi-confirmatory factor analysis (SCFA) model that addresses the specification of ``non-zero loadings'' through the covariance structure learned from high-dimensional data, and significantly alleviates the computational burden with theoretically guaranteed solutions in closed form.
	Specifically, to overcome the first limitation, we incorporate a prevalent covariance structure, namely, the \emph{interconnected community structure}, into the conventional CFA model. 
	We focus on the interconnected community structure, as it is widely prevalent in the covariance matrices of various high-dimensional datasets, as illustrated in Figure~\ref{Fig:ics_examples}.
	Notably, interconnected community structures, which enable features between communities to exhibit correlations, encompass various well-known patterns, including all independent community structures \citep{NewmanGirvan2004, Fortunato2010} and most hierarchical community structures \citep{LiLeiBhattacharyya2022, SchaubLiPeel2023}.
	Therefore, they provide a versatile covariance structure to model various practical applications, including brain imaging, gene expression, multi-omics, metabolomics, and more \citep{GirvanNewman2002, ColizzaFlamminiSerrano2006, SimpsonBowmanLaurienti2013, LevineSimondsBendall2015, HuttlinBrucknerPaulo2017, ZitnikSosicLeskovec2018, PerrotLevyRajjou2022}.
	However, we acknowledge that the proposed structures can be limited in representing certain covariance patterns (e.g., Toeplitz). 
	In such cases, traditional factor models remain suitable.  
	Interconnected community structures are latent in many studies, which can be accurately and robustly estimated and extracted by recently developed network structure detection approaches \citep{WangLiangJi2020, LiLeiBhattacharyya2022, YangChenChen2024}.
	Consequently, the detected community membership can serve as a guide to specifying the previously unknown ``non-zero loadings'' for CFA, effectively addressing the first limitation.
	The SCFA model also alleviates the computational burden by deriving closed-form solutions for all CFA model parameters, including the factor loadings, factors, covariance matrix between common factors, and covariance matrix for error terms.  
	The closed-form estimators not only improve estimation accuracy and stability but also drastically reduce computational load and ensure scalability (e.g., handling thousands of observed variables).
	
	SCFA presents several methodological contributions.
	Firstly, SCFA alleviates the requirement of ``non-zero loadings'' in CFA models. 
	The acquired interconnected community structure assigns observed variables to common factors, specifying ``non-zero loadings'' in an adaptive manner.
	Secondly, SCFA provides a computationally efficient approach for conducting high-dimensional CFA (e.g., with thousands or more observed variables).
	All estimators are obtained by the likelihood approach and in closed form, substantially mitigating the computational burden.
	Thirdly, SCFA yields more accurate and reliable estimates, since all matrix estimators are uniformly minimum-variance unbiased estimators (UMVUEs).
	The factor scores can also be conveniently estimated using the feasible generalized least-square (FGLS) method. 
	We further show that FGLS estimators have an identical solution to those obtained through ordinary least-square (OLS) and generalized least-square (GLS) methods.
	Lastly, we derive explicit variance estimators that facilitate statistical inference concerning model parameters and factor scores in a SCFA model.
	
	The remainder of the paper is structured as follows. 
	In Section~\ref{Sec:method}, we introduce the SCFA model, detailing its specifications for the factor loading matrix and the covariance matrix of observations.
	We subsequently present the estimation and inference procedures for all unknown matrices and factor scores. 
	Section~\ref{Sec:simulation} and Section~\ref{Sec:real} are dedicated to evaluating the proposed model and approach.
	Section~\ref{Sec:simulation} assesses the performance of our model through simulated data, while Section~\ref{Sec:real} demonstrates the applications to two genomics datasets without prior knowledge of ``non-zero loadings''.
	All proofs and additional tables are included in the \href{Supplementary Material.pdf}{Supplementary Material}.

\section{Method}
	
\label{Sec:method}
	
	\subsection{Background}
	
	\label{Subsec:background}
	
	\emph{Confirmatory factor analysis model.} 
	Let $\bm{X}_{p \times 1}$ represent a $p$-dimensional vector of observations, $\bm{f}_{K \times 1}$ denote the $K$-dimensional vector of common factors, with $1 < K < p$, and $\bm{u}_{p \times 1}$ denote the $p$-dimensional vector of error terms. 
	A factor model can be expressed as  
	\begin{equation}
		\label{Eq:factor_model}
		\bm{X}_{p \times 1} = \bm{\mu}_{p \times 1} + \textbf{L}_{p \times K} \bm{f}_{K \times 1} + \bm{u}_{p \times 1},
		\operatorname{E} (\bm{f}) = \textbf{0}_{K \times 1}, 
		\operatorname{E} (\bm{u}) = \textbf{0}_{p \times 1}, 
		\operatorname{cov} \left(\bm{f}, \bm{u} \right) = \textbf{0}_{K \times p},
	\end{equation}
	where $\textbf{0}_{p \times q}$ represents the $p$ by $q$ zero matrix; without loss of generality, the $p$-dimensional mean vector is denoted by $\bm{\mu} = \textbf{0}_{p \times 1}$; and $\textbf{L}$ represents the $p$ by $K$ factor loading matrix.
	Furthermore, let $\bm{\Sigma} \triangleq \operatorname{cov}(\bm{X})$, $\bm{\Sigma}_{\bm{f}} \triangleq \operatorname{cov}(\bm{f})$, and $\bm{\Sigma}_{\bm{u}} \triangleq \operatorname{cov}(\bm{u})$ denote the $p$ by $p$, $K$ by $K$, and $p$ by $p$ covariance matrices, respectively, where $\triangleq$ represents the equality by definition and $\bm{\Sigma}_{\bm{u}}$ is diagonal \citep{FanWangZhong2021}.
	
	When performing CFA, we may introduce zero entries at specified positions in the factor loading matrix $\textbf{L}$ \citep{Joreskog1969} and assume the common factors to be oblique, e.g., the covariance matrix of common factors $\bm{\Sigma}_{\bm{f}}$ can be arbitrarily positive definite.
	Following the model in~\eqref{Eq:factor_model} and these two assumptions of $\textbf{L}$ and $\bm{\Sigma}_{\bm{f}}$, we can derive the following relationship between covariance matrices in the CFA model:   
	\begin{equation}
		\label{Eq:covariance_model}
		\bm{\Sigma} 
		= \textbf{L} \bm{\Sigma}_{\bm{f}} \textbf{L} ^ \top + \bm{\Sigma}_{\bm{u}},
	\end{equation}
	where $\top$ denotes the transpose, $\textbf{L} = \operatorname{Bdiag}(\bm{\ell}_1, \ldots, \bm{\ell}_K)$ becomes a $p$ by $K$ block-diagonal matrix with $p_k$-dimensional ``non-zero loadings'' $\bm{\ell}_k$ for $k = 1, \ldots, K$; $\bm{\Sigma}_{\bm{f}} \triangleq \left(\sigma_{\bm{f}, kk'}\right)$ is a $K$ by $K$ symmetric positive-definite matrix; and $\bm{\Sigma}_{\bm{u}} \triangleq \left(\bm{\Sigma}_{\bm{u}, kk'}\right)$ is a $p$ by $p$ diagonal positive-definite matrix satisfying the submatrix $\bm{\Sigma}_{\bm{u}, kk'}$ is a $p_k$ by $p_k'$ diagonal matrix if $k' = k$ or zero matrix if $k' \neq k$.
	In particular, $p_k$ represents the number of observed variables within the $k$th factor, for $k = 1, \ldots, K$, satisfying that each $p_k > 1$ and $p = p_1 + \cdots + p_K$.
	
	We remark that the non-overlapping factor loading pattern in~\eqref{Eq:covariance_model} permits only a single non-zero entry within each row of $\textbf{L}$.
	This configuration aligns with the preferences of the majority of existing complexity criteria \citep{Browne2001}.
	
	\emph{Covariance matrix $\bm{\Sigma}$ in a block form.} 
	As defined in~\eqref{Eq:covariance_model}, the covariance matrix of observed variables $\bm{\Sigma}$ is structured as a block matrix. 
	Specifically, the block structure of $\bm{\Sigma} \triangleq \left(\bm{\Sigma}_{kk'}\right)$ is determined by the ``non-zero loadings'' $\bm{\ell}_k$ in $\textbf{L}$ for $k = 1, \ldots, K$: 
	\begin{equation*}
		\bm{\Sigma} \triangleq \left(\bm{\Sigma}_{kk'}\right)
		= \begin{pmatrix}
			\bm{\Sigma}_{11} & \bm{\Sigma}_{12} & \dots & \bm{\Sigma}_{1K} \\
			\bm{\Sigma}_{21} & \bm{\Sigma}_{22} & \dots & \bm{\Sigma}_{2K} \\
			\vdots & \vdots & \ddots & \vdots \\
			\bm{\Sigma}_{K1} & \bm{\Sigma}_{K2} & \dots & \bm{\Sigma}_{KK}
		\end{pmatrix}, \quad 
		\bm{\Sigma}_{kk'} = \sigma_{\bm{f}, kk'} \bm{\ell}_k \bm{\ell}_{k'} ^ \top + \bm{\Sigma}_{\bm{u}, kk'} \in \mathbb{R} ^ {p_k \times p_{k'}},
	\end{equation*}
	for all $k$ and $k'$, where $\bm{\Sigma}_{k'k} = \bm{\Sigma}_{kk'}$ for $k' \neq k$ and $\bm{\Sigma}_{\bm{u}, kk'}$ is a diagonal matrix.
	
	In practical applications, it is often the case that information about both the components and lengths of each ``non-zero loadings'' $\bm{\ell}_k$ is unavailable, particularly in domains such as omics and imaging data.
	Nevertheless, covariance matrices in these applications often exhibit block patterns, albeit these block patterns may be latent and require pattern extraction algorithms. 
	Recent advancements in statistics have provided convenient tools to accurately identify latent block structures in covariance matrices and precisely estimate covariance parameters \citep{LeiRinaldo2015, WuMaLiu2021, LiLeiBhattacharyya2022}.
	Inspired by the above block structure of $\bm{\Sigma} \triangleq \left(\bm{\Sigma}_{kk'}\right)$ in the CFA model, we are motivated to extract knowledge regarding all ``non-zero loadings'' from structured covariance matrix estimation for high-dimensional data, thereby facilitating the estimation of the CFA model, as elaborated in the following sections.

	\emph{Interconnected community structure.} 	
	The block-structured covariance matrix in the CFA model is naturally linked with the community-based network structure \citep{WuMaLiu2021}. 
	In the current research, we focus on the interconnected community structure, which is more general and prevalent in real-world applications. 
	As demonstrated in Figure~\ref{Fig:ics_examples}, the interconnected community covariance structure presents in the high-throughput datasets such as genetics, imaging, gene-expression, DNA-methylation, metabolomics, and finance, among many \citep{ChenFanZhu2023, YangChenChen2024}.
	Therefore, we build the proposed model based on the interconnected community covariance structure for factor analysis on these datasets.
	
	We characterize the interconnected community covariance structure as follows. 
	In this structure, all features can be categorized into multiple mutually exclusive and exhaustive communities. 
	In other words, there implicitly exists a community-membership function $\varphi$ that operates as a bijection.
	In contrast to the classical community network structure, the interconnected community structure exhibits correlations among features within and between communities (i.e., interactive communities), as demonstrated in Figure~\ref{Fig:ics_examples}.
	Thus, the (population) covariance matrix with an interconnected community structure has a block structure: diagonal blocks represent the intra-community correlations while the off-diagonal blocks characterize the inter-community relationships. 
	Due to the high resemblance of entries within each block, a parametric covariance model is often used by assigning two (or one) parameters for each diagonal (or off-diagonal) block \citep{YangChenChen2024}.
	
	A two-step procedure is commonly employed for estimating large covariance matrices in these datasets: firstly, extracting the latent structure from the sample covariance matrix, and subsequently estimating the covariance parameters under the learned covariance structure \citep{ChenKangXing2018, WuMaLiu2021, ChenZhangWu2024}.
	Given the reliable and replicable performance of existing algorithms in extracting latent network structures \citep{ChenKangXing2018, WangLiangJi2020, LiLeiBhattacharyya2022} (see almost identical results of detected interconnected structures by different algorithms in the \href{Supplementary Material.pdf}{Supplementary Material}), it is plausible to consider the estimated interconnected community structure of the covariance matrix as \emph{known} prior knowledge for SCFA models.
	
	\begin{figure}[!htb]
		\centering
		\includegraphics[width = 1 \linewidth]{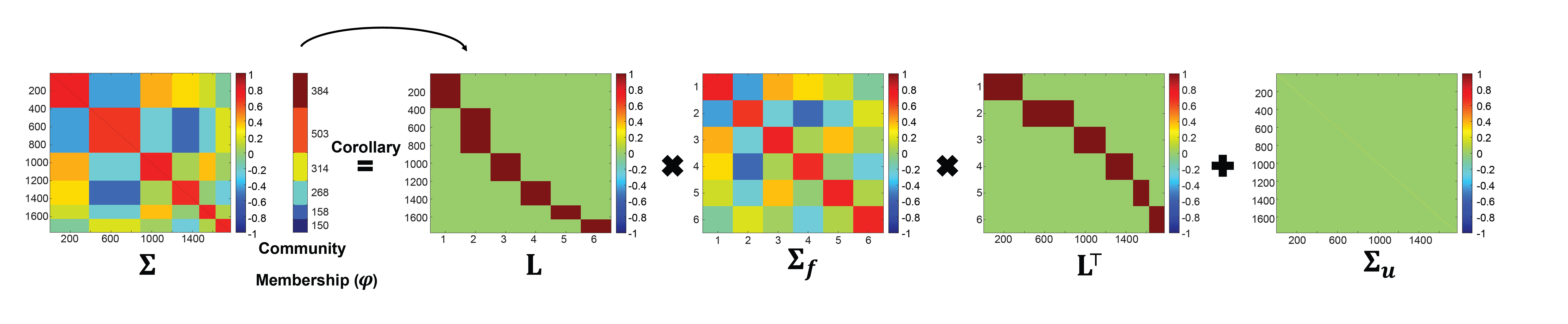} 
		\caption{
			The overview of SCFA from the perspective of a covariance matrix. 
			The left subfigure illustrates the population covariance matrix $\bm{\Sigma}$ with an explicit interconnected community structure. 
			The parameters of the SCFA model, $\textbf{L}$, $\bm{\Sigma}_{\bm{f}}$, and $\bm{\Sigma}_{\bm{u}}$, are linked with $\bm{\Sigma}$ as demonstrated by the right-hand side of the equation. 
			The community membership in the interconnected community structure can specify the ``non-zero loadings'' in $\textbf{L}$, while $\bm{\Sigma}_{\bm{f}}$ and $\bm{\Sigma}_{\bm{u}}$ are determined by the parametric covariance structure in $\bm{\Sigma}$.
		}
		\label{Fig:decomposition}
	\end{figure}
	
	\subsection{Semi-parametric confirmatory analysis model}
	
	\label{Subsec:scfa}
	
	We propose a semi-confirmatory factor analysis (SCFA) model for multivariate $X$ with a covariance matrix $\Sigma$ having an interconnected community structure, as illustrated in Figure~\ref{Fig:decomposition}. 
	SCFA specifies non-zero loadings based on the learned interconnected community structure without the requirement of prior knowledge.
	Without loss of generality, we further assume that the covariance matrix $\bm{\Sigma}$ has a parametric representation (see matrix $\bm{\Sigma}_{p \times p}$ in~\eqref{Eq:scfa} and details in \citet{YangChenChen2024}). 
	Specifically, following~\eqref{Eq:factor_model} and~\eqref{Eq:covariance_model}, we present the SCFA model as:
	\begin{gather}
		\bm{X} = \textbf{L} \bm{f} + \bm{u}, \quad \bm{\Sigma} = \textbf{L} \bm{\Sigma}_{\bm{f}} \textbf{L} ^ \top + \bm{\Sigma}_{\bm{u}}, \text{ where } \label{Eq:scfa} \\
		\begin{align*}
			\textbf{L}_{p \times K} 
			& = \operatorname{Bdiag} \left(\bm{\ell}_1, \ldots, \bm{\ell}_K\right) 
			\text{ with } \bm{\ell}_k \triangleq \begin{pmatrix}
				\ell_{\bar{p}_{k - 1} + 1, k} \\ \vdots \\ \ell_{\bar{p}_{k}, k}
			\end{pmatrix}_{p_k \times 1},  
			\ell_{\bar{p}_{k - 1} + 1, k} \triangleq \tau_k \neq 0,  \\
			\bm{\Sigma}_{p \times p} 
			& \triangleq \left(\bm{\Sigma}_{kk'}\right) 
			\triangleq \begin{pmatrix}
				a_{11} \textbf{I}_{p_1} + b_{11} \textbf{J}_{p_1} & b_{12} \textbf{1}_{p_1 \times p_2} & \cdots & b_{1K} \textbf{1}_{p_1 \times p_K} \\
				b_{21} \textbf{1}_{p_2 \times p_1} & a_{22} \textbf{I}_{p_2} + b_{22} \textbf{J}_{p_2} & \cdots & b_{2K} \textbf{1}_{p_2 \times p_K} \\ 
				\vdots & \vdots & \ddots & \vdots \\
				b_{K1} \textbf{1}_{p_K \times p_1} & b_{K2} \textbf{1}_{p_K \times p_2} & \cdots & a_{KK} \textbf{I}_{p_K} + b_{KK} \textbf{J}_{p_K} 
			\end{pmatrix}, b_{kk'} = b_{k'k},
		\end{align*}
	\end{gather}
	where $k, k' = 1, \ldots, K$ and $K$, the number of common factors, is set to be the number of interconnected communities.
	$p_1, \ldots, p_K$ are set to be the interconnected community sizes satisfying that $p_k > 1$ for every $k$ and $p = p_1 + \cdots + p_K$, and $\bar{p}_k = \sum_{k' = 1}^{k} p_{k'}$ is the sum (we define $\bar{p}_0 = 0$). 
	Without loss of generality, the first entry in $\bm{\ell}_k$ is assumed to be a \emph{known} non-zero constant, i.e., $\ell_{\bar{p}_{k - 1} + 1, k} \triangleq \tau_k \neq 0$ is known, while the other entries in $\bm{\ell}_k$ are \emph{unknown} for every $k$.
	$\textbf{I}_p$ denotes the $p$ by $p$ identity matrix, $\textbf{J}_p$ and $\textbf{1}_{p \times q}$ denote $p$ by $p$ and $p$ by $q$ all-one matrices, respectively. 
	$a_{kk}$ and $b_{kk'}$ are \emph{unknown} entries of the parametric population covariance matrix.
	
	The parametric covariance matrix $\bm{\Sigma}$ defined in~\eqref{Eq:scfa} is derived from covariance patterns inherent in interconnected community structures \citep{YangChenChen2024}.
	When viewed through the lens of network analysis, $\bm{\Sigma}$ takes a block form, where each diagonal block represents a uniform correlation relationship between features within a single community, and each off-diagonal block represents a uniform correlation relationship between features from two distinct communities.
	We show that the SFCA model is generally robust to the misspecification of the parametric covariance matrix in the simulation section. 
	
	Under this specification, the SCFA model is distinct from the conventional CFA model because it can more precisely allocate correlated variables to common factors, more flexibly model the covariance matrix between common factors, and provide closed-form parameter estimates for $\textbf{L}$, $\bm{\Sigma}_{\bm{f}}$, and $\bm{\Sigma}_{\bm{u}}$. 
	
	
	\emph{Specifying all ``non-zero loadings'' in $\textbf{L}$.}	
	In the SCFA model, we specify the ``non-zero loadings'' in $\textbf{L}$ based on the interconnected community structure. 
	For an interconnected community structure of $K$ communities with corresponding community sizes $\bm{\ell}_1, \ldots, \bm{\ell}_K$, the factor loading matrix is
	\begin{equation*}
		\textbf{L} 
		= \begin{pmatrix}
			\ell_{\bar{p}_{0} + 1, 1} &  & & \\
			\ell_{\bar{p}_{0} + 2, 1} & & & \\
			\vdots 					  & & & \\
			\ell_{\bar{p}_{1}, 1}     & & & \\
			& \ell_{\bar{p}_{1} + 1, 2}		& 		 & \\
			& \ell_{\bar{p}_{1} + 2, 2}		&		 & \\
			& \vdots & 		 & \\
			& \ell_{\bar{p}_{2}, 2} 		&		 & \\
			& 		& \ddots & \\
			&		& 		 & \ell_{\bar{p}_{K - 1} + 1, K} \\ 
			&		& 		 & \ell_{\bar{p}_{K - 1} + 2, K} \\
			&		& 		 & \vdots \\
			&		& 		 & \ell_{\bar{p}_{K}, K} \\
		\end{pmatrix}_{p \times K}.
	\end{equation*}
	
	We utilize the community-membership function $\varphi$ based on the interconnected community structure to partition $p$ observed variables into $K$ mutually exclusive and exhaustive common factors, i.e., $\varphi: \{1, 2, \ldots, p \} \to \{1, 2, \ldots, K \}, j \mapsto \varphi(j)$.
	Without loss of generality, the $j$th observed variable is mapped to the $k$th common factor, which includes $p_k$ distinct observed variables $\{j: \varphi(j) = k \} = \{\varphi ^ {-1}(k)[1], \ldots, \varphi ^ {-1}(k)[p_k]\}$ and satisfies that $p = p_1 + \cdots + p_K$.
	Next, we reorder $p$ observed variables by listing them from the same common factor as neighbors.
	For the input vector of observed variables $\bm{X} \triangleq \left(X ^ {(1)}, \ldots, X ^ {(p)} \right)^\top$ in~\eqref{Eq:factor_model}, we reorder the elements and obtain $\bm{X} ^ {(\varphi)} = \left (X ^ {\left(\varphi^{-1}(1)[1]\right)}, \ldots, X ^ {\left(\varphi^{-1}(1)[p_1]\right)}, \ldots, X ^ {\left(\varphi^{-1}(K)[1]\right)}, \ldots, X ^ {\left(\varphi^{-1}(K)[p_K]\right)} \right) ^ \top$ using the function $\varphi$.
	Consequently, $\varphi$ dictates that the first $p_1$ elements of $\bm{X} ^ {(\varphi)}$ are categorized into the first common factor, and so forth, with the last $p_K$ elements of $\bm{X} ^ {(\varphi)}$ being categorized into the $K$th common factor.
	For simplicity, we denote $\bm{X} ^ {(\varphi)} = \bm{X}$ throughout the remainder of this paper. 
	Moreover, the adoption of a non-overlapping factor loading pattern in~\eqref{Eq:covariance_model} and~\eqref{Eq:scfa} stems from the fact that $\varphi$ is a bijective function.
	
	With the exception of components $\ell_{\bar{p}_k + j, k + 1}$ for $j = 2, \ldots, p_{k + 1}$ and $k = 1, \ldots, K$, which will be determined in the subsequent lemma, all information about the factor loading matrix has been completely determined by the community membership.

	\emph{Reparameterizing $\bm{\Sigma}_{\bm{f}}$ and $\bm{\Sigma}_{\bm{u}}$ by $\bm{\Sigma}$.} 
	With a covariance matrix $\bm{\Sigma}$ characterized by an interconnected community structure and corresponding ``non-zero loadings'' specified in $\textbf{L}$, the SCFA model in~\eqref{Eq:scfa} can represent model parameters $\bm{\Sigma}_{\bm{f}}$ and $\bm{\Sigma}_{\bm{u}}$ using $K + (K+1)K / 2$ covariance parameters in $\bm{\Sigma}$.
	
	\begin{lemma}
		\label{Lem:decomposition}
		Consider the block structure of the covariance matrix in a classical CFA model as shown in~\eqref{Eq:covariance_model} with each submatrix $\bm{\Sigma}_{kk'} = \sigma_{\bm{f}, kk'} \bm{\ell}_k \bm{\ell}_{k'} ^ \top + \bm{\Sigma}_{\bm{u}, kk'}$.
		By SCFA, all parameters in each $\bm{\Sigma}_{kk'}$ can be determined by $a_{kk}$ and $b_{kk'}$ in~\eqref{Eq:scfa} through: 
		\begin{equation*}
			\bm{\Sigma}_{kk'}
			= \sigma_{\bm{f}, kk'} \bm{\ell}_k \bm{\ell}_{k'} ^ \top + \bm{\Sigma}_{\bm{u}, kk'}
			= \begin{cases}
				a_{kk} \textbf{I}_{p_k} + b_{kk} \textbf{J}_{p_k}, & k' = k \\
				b_{kk'} \textbf{1}_{p_k \times p_{k'}}, & k' \neq k
			\end{cases}
			, b_{kk'} = b_{k'k},
		\end{equation*}
		with $p_k > 2$ for every $k$.
		Then, we have the following equations:
		
		(1) $\bm{\ell}_k \triangleq \begin{pmatrix}
			\ell_{\bar{p}_{k - 1} + 1, k} \\ \vdots \\ \ell_{\bar{p}_{k}, k}
		\end{pmatrix} = \begin{pmatrix}
			\tau_k \\ \vdots \\ \tau_k
		\end{pmatrix} = \tau_k \textbf{1}_{p_k \times 1}$ for all $k$, so 
		$\textbf{L} = \operatorname{Bdiag} \left(\tau_1 \textbf{1}_{p_1 \times 1}, \ldots, \tau_K \textbf{1}_{p_K \times 1}\right)$;
		
		(2) $\sigma_{\bm{f}, kk'} = \dfrac{b_{kk'}}{\tau_k \tau_{k'}}$ for all $k, k'$, so $\bm{\Sigma}_{\bm{f}} \triangleq \left(\sigma_{\bm{f}, kk'}\right) = \left(\dfrac{b_{kk'}}{\tau_k \tau_{k'}}\right)$  with $b_{kk'} = b_{k'k}$;
		
		(3) $\bm{\Sigma}_{\bm{u}, kk} = a_{kk} \textbf{I}_{p_k}$ for all $k$, so $\bm{\Sigma}_{\bm{u}} \triangleq \left(\bm{\Sigma}_{\bm{u}, kk}\right) = \operatorname{Bdiag}\left(a_{11} \textbf{I}_{p_1}, \ldots, a_{KK}  \textbf{I}_{p_K}\right)$, 
		where we assume that the symmetric matrix $\left(b_{kk'}\right)$ is positive definite and $a_{kk} > 0$ for all $k$.
	\end{lemma}
	
	The conditions $a_{kk} > 0$ for all $k$ and the positive definiteness of $\left(b_{kk'}\right)$ are required to ensure that $\bm{\Sigma}$ is positive definite (see \href{Supplementary Material.pdf}{Supplementary Material}).
	Lemma~\ref{Lem:decomposition} demonstrates that the SCFA model can (1) directly specify the ``non-zero loadings'' in $\textbf{L}$ based on the function $\varphi$ and derive that all loadings belonging to factor $k$ follow $\ell_{\bar{p}_{k-1} + 1, k} = \cdots = \ell_{\bar{p}_{k}, k} = \tau_k \neq 0$; and (2) parameters for all entries in the matrices $\bm{\Sigma}_{\bm{f}}$ and $\bm{\Sigma}_{\bm{u}}$ can be expressed by the parameters in the parametric covariance matrix $\bm{\Sigma}$ (i.e., $a_{kk}$ and $b_{kk'}$). 
	This reparametrization facilitates much improved parameter estimation accuracy and computational efficiency for SCFA.
	
	\subsection{Estimation and inference}
	
	\label{Subsec:estimation_inference}
	
	In this section, we introduce a likelihood-based parameter estimation procedure for SCFA applied to sample data $\textbf{X}_{n \times p}$ with an unknown covariance matrix having an interconnected community structure.
	We derive closed-form estimators for SCFA parameters $\textbf{L}$, $\bm{\Sigma}_{\bm{f}}$ and $\bm{\Sigma}_{\bm{u}}$, and calculate closed-form factor scores using the least-square approach. 
	Lastly, we establish the theoretical properties of the proposed estimators and delineate the inference procedure.
	
	Suppose that the rows $\bm{X}_1, \ldots, \bm{X}_n$ of data matrix $\textbf{X}_{n \times p}$ are independently and identically distributed as $N \left(\textbf{0}_{p \times 1}, \bm{\Sigma} \right)$, satisfying the proposed SCFA model in~\eqref{Eq:scfa}.
	Let $\bm{f}_i$ and $\bm{u}_i$ be the $i$th factor score and error term, respectively.
	We define $\textbf{S}_{p \times p} = n ^ {-1} \textbf{X} ^ \top \textbf{X}$ as the unbiased sample covariance matrix.
	The typical estimation objectives encompass the $(p - K)$ non-zero elements of $\textbf{L}$, the $K(K + 1)/2$ elements of symmetric $\bm{\Sigma}_{\bm{f}}$, and the $p$ diagonal elements of $\bm{\Sigma}_{\bm{u}}$ based on $\textbf{S}$. 
	We assume that the interconnected community structure and $\varphi$ are known for $\bm{\Sigma}$, while parameters $a_{kk}$ and $b_{kk'}$ need to be estimated for each $k$ and $k'$.
	Without loss of generality, we let the high-dimensional framework in this paper follow that $n < p$ and $K + K(K + 1) / 2 < n$.
	
	We maximize the likelihood function with respect to $\textbf{L} = \operatorname{Bdiag}\left(\bm{\ell}_{1}, \ldots, \bm{\ell}_{K} \right)$, $\bm{\Sigma}_{\bm{f}} = \left(\dfrac{b_{kk'}}{\tau_k \tau_{k'}}\right)$, and $\bm{\Sigma}_{\bm{u}} = \operatorname{Bdiag}\left(a_{11} \textbf{I}_{p_1}, \ldots, a_{KK} \textbf{I}_{p_K} \right)$:
	\begin{equation*}
		\begin{split}
			\left(\widehat{\textbf{L}}, \widehat{\bm{\Sigma}}_{\bm{f}}, \widehat{\bm{\Sigma}}_{\bm{u}}\right)
			& = \argmax \limits_{\textbf{L}, \bm{\Sigma}_{\bm{f}}, \bm{\Sigma}_{\bm{u}}} \mathcal{L}_n \left(\textbf{L}, \bm{\Sigma}_{\bm{f}}, \bm{\Sigma}_{\bm{u}}; \textbf{S}, \tau_1, \ldots, \tau_K \right) \\
			& = \argmax \limits_{a_{11}, \ldots, a_{KK}, b_{11}, b_{12}, \ldots, b_{KK}} \mathcal{L}_n \left(a_{11}, \ldots, a_{KK}, b_{11}, b_{12}, \ldots, b_{KK}, \widehat{\textbf{L}}; \textbf{S}, \tau_1, \ldots, \tau_K \right) \\
			& = \argmax \limits_{a_{11}, \ldots, a_{KK}, b_{11}, b_{12}, \ldots, b_{KK}} \bigg [ - \frac{n}{2} \log \det \left(\bm{\Sigma} \left(a_{11}, \ldots, a_{KK}, b_{11}, b_{12}, \ldots, b_{KK}\right)\right) \\
			& - \frac{n}{2} \operatorname{tr} \left\{\textbf{S} \bm{\Sigma} ^ {- 1} \left(a_{11}, \ldots, a_{KK}, b_{11}, b_{12}, \ldots, b_{KK}\right) \right\} \bigg ],
		\end{split} 
	\end{equation*}
	where $\mathcal{L}_n$ denotes the log-likelihood function of normal data $\textbf{X}$, $\det(\cdot)$ denotes the determinant, $\operatorname{tr}(\cdot)$ denotes the trace, $\bm{\Sigma} \triangleq \bm{\Sigma} \left(a_{11}, \ldots, a_{KK}, b_{11}, b_{12}, \ldots, b_{KK}\right)$ is defined in~\eqref{Eq:scfa}. 
	By maximizing $\mathcal{L}_n$ and letting $\tau_1 = \cdots = \tau_K = 1$ throughout the rest of the paper, we obtain the following (unique) maximum likelihood estimators in closed forms: 
	\begin{equation*}
		\widehat{\textbf{L}} = \operatorname{Bdiag} \left(\textbf{1}_{p_1 \times 1}, \ldots, \textbf{1}_{p_K \times 1}\right), \quad
		\widehat{\bm{\Sigma}}_{\bm{u}} = \operatorname{Bdiag}\left(\widehat{a}_{11} \textbf{I}_{p_1}, \ldots, \widehat{a}_{KK} \textbf{I}_{p_K}\right), \quad
		\widehat{\bm{\Sigma}}_{\bm{f}} = \left(\widehat{b}_{kk'}\right),
	\end{equation*}
	with $\widehat{b}_{kk'} = \widehat{b}_{k'k}$,
	where
	\begin{equation}
		\label{Eq:mle}
		\widehat{a}_{kk} = \frac{p_k \operatorname{tr}\left(\textbf{S}_{kk}\right) - \operatorname{sum}\left(\textbf{S}_{kk}\right)}{p_k (p_k - 1)}, \quad
		\widehat{b}_{kk'} = \begin{cases}
			\dfrac{\operatorname{sum}\left(\textbf{S}_{kk'}\right)}{p_k p_{k'}}, & k \neq k' \\
			\dfrac{\operatorname{sum}\left(\textbf{S}_{kk}\right) - \operatorname{tr}\left(\textbf{S}_{kk}\right)}{p_k (p_k - 1)}, & k = k'
		\end{cases},
	\end{equation}
	for all $k$ and $k'$, $\textbf{S}_{kk'}$ denotes the $p_k$ by $p_{k'}$ submatrix of $\textbf{S} \triangleq \left(\textbf{S}_{kk'}\right)$, $\operatorname{sum}(\cdot)$ denotes the sum of all elements of a matrix.
	
	The estimation procedure for SCFA parameters is scalable for high-dimensional data (e.g., $p > 10^4$) as long as $n > K + K (K + 1) / 2$. 
	In contrast, the classic CFA model faces challenges in computation when (1) $p > n$ and (2) $n > p$ but $p$ is several hundred, due to the bottleneck of computing the large covariance (and its inverse) using the maximum likelihood approach.
	Therefore, SCFA addresses a longstanding limitation in  CFA regarding dimensionality constraints.
	
	Furthermore, the proposed factor score estimator of $\bm{f}_i$ is
	\begin{equation}
		\label{Eq:factor_scores} 
		\widehat{\bm{f}}_i
		= \left(\widehat{\textbf{L}} ^ \top \widehat{\textbf{L}}\right) ^ {- 1} \widehat{\textbf{L}} ^ \top \bm{X}_i 
		= \left(\widehat{\textbf{L}} ^ \top \bm{\Sigma}_{\bm{u}} ^ {-1} \widehat{\textbf{L}}\right) ^ {- 1} \widehat{\textbf{L}} ^ \top \bm{\Sigma}_{\bm{u}} ^ {-1} \bm{X}_i
		= \left(\widehat{\textbf{L}} ^ \top \widehat{\bm{\Sigma}}_{\bm{u}} ^ {-1} \widehat{\textbf{L}}\right) ^ {- 1} \widehat{\textbf{L}} ^ \top \widehat{\bm{\Sigma}}_{\bm{u}} ^ {-1} \bm{X}_i, 
	\end{equation}
	for $i = 1, \ldots, n$, where the OLS estimator is identical to the generalized least-square (GLS) estimator and the feasible generalized least-square (FGLS) estimator.
	The derivation is provided in the \href{Supplementary Material.pdf}{Supplementary Material}.
	The following theorems exhibit the theoretical properties of the above estimators.
	
	\begin{theorem}
		\label{Thm:estimators}
		If $K + K(K + 1) / 2 < n$, 
		then the proposed matrix estimators $\widehat{\textbf{L}}$, $\widehat{\bm{\Sigma}}_{\bm{f}}$, and $\widehat{\bm{\Sigma}}_{\bm{u}}$ are uniformly minimum-variance unbiased estimators (UMVUEs). 
	\end{theorem}
	
	Please refer to the proof in the \href{Supplementary Material.pdf}{Supplementary Material}.
	As $\widehat{a}_{kk}$ and $\widehat{b}_{kk'}$ are (unique) maximum likelihood estimators, they also exhibit large-sample properties such as consistency, asymptotic efficiency, and asymptotic normality, under the conditions that $K + K(K + 1) / 2 < n$ and $n \to \infty$ for fixed $K$ and $p$.
	
	\begin{theorem}
		\label{Thm:factor_scores}
		If $K + K(K + 1) / 2 < n$ and $\bm{\Sigma}_{\bm{u}}$ is positive definite (i.e., $a_{kk} > 0$ for every $k$), 
		then $\widehat{\bm{f}}_i$ defined as~\eqref{Eq:factor_scores} is UMVUE, 
		following a multivariate normal distribution with mean $\bm{f}_i$ and covariance matrix presented in Theorem~\ref{Thm:vars}. 
	\end{theorem}
	
	The proofs of the equivalence~\eqref{Eq:factor_scores} and Theorem~\ref{Thm:factor_scores} are elaborated in the \href{Supplementary Material.pdf}{Supplementary Material}. 
	$\widehat{\bm{f}}_i$ also exhibits large-sample properties such as consistency, asymptotic efficiency, and asymptotic normality, as $p \to \infty$.
	The normality result in Theorem~\ref{Thm:factor_scores} extends the confidence intervals for factor scores $\bm{f}_i$.  
	In addition to the maximum likelihood estimators $\widehat{a}_{kk}$ and $\widehat{b}_{kk'}$, we provide their exact closed-form variance estimators in the following theorem.
	Moreover, the exact covariance matrix of the proposed factor score estimator $\widehat{\bm{f}}_i$ is~\eqref{Eq:factor_scores}.
	
	\begin{theorem}
		\label{Thm:vars}
		
		(1) The exact variance estimators of $\widehat{a}_{kk}$ and $\widehat{b}_{kk'}$ are 
		\begin{equation*}
			\begin{split}
				\operatorname{var} (\widehat{a}_{kk})
				& = \frac{2 a_{kk} ^ 2}{(n - 1)(p_k - 1)}, \\
				\operatorname{var} (\widehat{b}_{kk'})
				& = \begin{cases}
					\frac{2}{(n - 1)p_k(p_k - 1)} \left\{
					(a_{kk} + p_{k} b_{kk}) ^ 2 - (2 a_{kk} + p_k b_{kk}) b_{kk}
					\right\}, & k = k' \\
					\frac{1}{2 (n - 1) p_k p_{k'}} \left\{ p_k p_{k'} (b_{kk'} ^ 2 + b_{k'k} ^ 2) + 2 (a_{kk} + p_k b_{kk})(a_{k'k'} + p_{k'} b_{k'k'}) \right\}, & k \neq k'
				\end{cases}
			\end{split}
		\end{equation*}
		for every $k$ and $k'$.
		
		(2) The exact covariance matrix estimator of $\widehat{\bm{f}}_i$ is
		\begin{equation*}
			\begin{split}
				\operatorname{cov}\left(\widehat{\bm{f}}_i\right)
				= \left(\widehat{\textbf{L}} ^ \top \widehat{\textbf{L}}\right) ^ {- 1} \widehat{\textbf{L}} ^ \top
				\bm{\Sigma}
				\widehat{\textbf{L}} \left(\widehat{\textbf{L}} ^ \top \widehat{\textbf{L}}\right) ^ {- 1} 
				= \operatorname{diag} \left(\frac{a_{11}}{p_1}, \ldots, \frac{a_{KK}}{p_K}\right) + \left(b_{kk'}\right).
			\end{split}
		\end{equation*}
		In particular, as $p_k \to \infty$ for all $k$, then $p \to \infty$ and $\operatorname{cov}\left(\widehat{\bm{f}}_i\right) \to \left(b_{kk'}\right) = \bm{\Sigma}_{\bm{f}}$ for each $i$.
	\end{theorem}
	
	Utilizing Theorem \ref{Thm:vars} and the estimates provided in~\eqref{Eq:mle}, we can perform Wald-type hypothesis tests and compute interval estimates for all model parameters $\bm{\Sigma}_{\bm{f}}$, $\bm{\Sigma}_{\bm{u}}$, and $\bm{f}_i$.
	
\section{Simulation Studies}
	
\label{Sec:simulation}
	
	\subsection{Data generation} 
	
	\label{Subsec:generation}
	
	We perform Monte Carlo simulation to generate the observation vector $\bm{X}_i = \textbf{L} \bm{f}_i + \bm{u}_i$ for $i = 1, \ldots, n$, where the factor loading matrix $\textbf{L} = \operatorname{Bdiag}\left(\textbf{1}_{p_1 \times 1}, \ldots, \textbf{1}_{p_K \times 1}\right)$, the common factor $\bm{f}_i \sim N \left(\textbf{0}_{K \times 1}, \left(b_{kk'}\right)\right)$, and the error term $\bm{u}_i \sim N \left(\textbf{0}_{p \times 1}, \operatorname{Bdiag}(a_{11} \textbf{I}_{p_1}, \ldots, a_{KK} \textbf{I}_{p_K})\right)$ for $i = 1, \ldots, n$.
	Specifically, we explore various values for $(p_1, \ldots, p_K) ^ \top$ (as indicated in Table~\ref{Tab:study_2}) under different sample sizes, i.e., $n \in \{40, 80, 120\}$, with $K = 3$ and
	\begin{equation*}
		a_{11} = 0.1, a_{22} = 0.2, a_{33} = 0.5, \quad
		\left(b_{kk'}\right) = \begin{pmatrix}
			2.02 & 0.73 & 1.15 \\
			& 3.13 & 1.63 \\
			&  & 3.69
		\end{pmatrix} \text{ is symmetric}.
	\end{equation*}
	We repeat the above data generation procedure $100$ times. 
	
	\subsection{Assessing the estimated factor scores and model parameters}
	
	\label{Subsec:study_12}
	
	We apply the SCFA model to each simulated dataset to estimate $\textbf{L}$, $\bm{f}_i$, $\bm{\Sigma}_{\bm{f}}$, and $\bm{\Sigma}_{\bm{u}}$, respectively.
	Since a primary objective of CFA is dimension reduction, yielding reliable common factors, we first focus on evaluating the performance of $\bm{f}_i$ estimation. 
	Specifically, we employ the Euclidean loss  $\sum_{i=1}^{n}\left \Vert \widehat{\bm{f}}_i - \bm{f}_i \right \Vert$ as the evaluation criterion. 
	We calculate $\widehat{\bm{f}}_i$ by~\eqref{Eq:factor_scores} and benchmark it against competing methods including the various CFA computational methods implemented by the R packages of ``sem'' \citep{Fox2006, RFoxNieByrnes2022}, ``OpenMx'' \citep{NealeHunterPritikin2016, RBokerNealeMaes2023}, and ``lavaan'' \citep{Rosseel2012, RRosseelJorgensenRockwood2023}, respectively. 
	We constrain the scales of loadings by all methods to be identical for fair comparison.
	
	In Table~\ref{Tab:study_1}, we summarize the mean and standard deviation of the Euclidean losses for each approach across the $100$ replicates under different settings.
	The results indicate that SCFA outperforms conventional numerical approaches, exhibiting the lowest average loss, while all other methods demonstrate, on average, twice the SCFA loss. 
	Additionally, SCFA shows a much-reduced variation in loss (e.g., around one-fifth) compared to the competing methods. 
	Lastly, SCFA drastically improves computational efficiency by at least $100$ times. 
	As mentioned earlier, the numerical implementations of the conventional CFA method may be limited for input datasets with $p \geq n$ and may thus yield no results (i.e., ``NA'' entries in Table~\ref{Tab:study_1}). 
	In contrast, SCFA provides a viable solution for datasets with $p \geq n$, which are very common in practice (e.g., omics, imaging, and financial data).

	\begin{table}[!htb]
		\centering
		\caption{Results of the means, standard deviations (s.d.), and computation times (in seconds) for the Euclidean losses $\sum_{i = 1}^{n} \Vert \widehat{\bm{f}}_i - \bm{f}_i \Vert$ obtained from our proposed method and computational packages across $100$ replicates, given different values of $p$ and $n$, where ``NA'' indicates ``not available''.}
		\tiny	
		\begin{tabular}{cccccccccccccccc}
			\hline
			&  & SCFA &  &  &  & sem &  &  &  & OpenMx &  &  &  & lavaan &  \\ \cline{2-4} \cline{6-8} \cline{10-12} \cline{14-16} 
			$(n, p)$ & mean & s.d. & time &  & mean & s.d. & time &  & mean & s.d. & time &  & mean & s.d. & time \\ \hline
			$(40, 20)$ & 12.16 & 0.78 & 0.04 &  & 20.7 & 5.93 & 23.03 &  & 21.31 & 5.84 & 139.89 &  & 21.24 & 5.84 & 16.21 \\
			$(40, 30)$ & 10.03 & 0.67 & 0.14 &  & 20.26 & 6.95 & 102.29 &  & 20.7 & 6.85 & 331.37 &  & 20.66 & 6.86 & 22.39 \\
			$(40, 40)$ & 8.72 & 0.58 & 0.04 &  & NA & NA & NA &  & 19.53 & 8.07 & 866.16 &  & NA & NA & NA \\
			$(80, 40)$ & 17.26 & 0.87 & 0.04 &  & 30.06 & 10.06 & 423.17 &  & 30.98 & 9.75 & 904.6 &  & 30.93 & 9.79 & 47.75 \\
			$(80, 80)$ & 12.27 & 0.65 & 0.05 &  & NA & NA & NA &  & 31.05 & 14.22 & 262.32 &  & NA & NA & NA \\
			$(80, 120)$ & 9.83 & 0.49 & 0.22 &  & NA & NA & NA &  & NA & NA & NA &  & NA & NA & NA \\
			$(120, 40)$ & 25.88 & 1.01 & 0.04 &  & 39.75 & 10.98 & 409.12 &  & 40.58 & 10.82 & 866.05 &  & 40.54 & 10.82 & 47.29 \\
			$(120, 120)$ & 14.95 & 0.6 & 0.07 &  & NA & NA & NA &  & NA & NA & NA &  & NA & NA & NA \\
			$(120, 200)$ & 11.49 & 0.51 & 0.13 &  & NA & NA & NA &  & NA & NA & NA &  & NA & NA & NA \\ \hline
		\end{tabular}
		\label{Tab:study_1}
	\end{table}
	
	In addition to the factor scores, we evaluate the performance of estimating model parameters $\bm{\Sigma}_{\bm{f}}$ and $\bm{\Sigma}_{\bm{u}}$. 
	As parametric matrices $\bm{\Sigma}_{\bm{f}}$ and $\bm{\Sigma}_{\bm{u}}$ can be represented by parameters $a_{kk}$ and $b_{kk'}$, we assess the accuracy of the estimators $\widehat{a}_{kk}$ and $\widehat{b}_{kk'}$ in~\eqref{Eq:mle}.
	
	The estimation results are summarized in Table~\ref{Tab:study_2}. 
	We consider the following metrics in Table~\ref{Tab:study_2}: the average bias, Monte Carlo standard deviation, average standard error, and $95\%$ Wald-type empirical coverage probability using the proposed estimates for each $a_{kk}$ and $b_{kk'}$. 
	The results in Table~\ref{Tab:study_2} demonstrate that our estimation is generally accurate with small biases and approximate $95\%$ coverage probabilities.
	Specifically, for each parameter, the bias is relatively small when compared to the Monte Carlo standard deviation, while the average standard error is close to the Monte Carlo standard deviation.
	Furthermore, both the bias and average standard error decrease as the sample size increases, and the $95\%$ coverage probability approaches the nominal level as the sample size grows.
	
	In comparison, we also utilize the aforementioned three R packages to estimate all $\bm{\ell}_k$ in the factor loading matrix $\textbf{L}$, all diagonal elements of $\bm{\Sigma}_{\bm{u}} = \operatorname{Bdiag}\left(a_{11} \textbf{I}_{p_1}, \ldots, a_{KK} \textbf{I}_{p_K} \right)$, and all elements of $\bm{\Sigma}_{\bm{f}} = \left(b_{kk'}\right)$.
	We also calculate the average bias and asymptotic standard error using the results produced by ``sem'', ``OpenMx'', and ``lavaan'', respectively, for all elements of $\bm{\ell}_k$, $\bm{\Sigma}_{\bm{u}} = \operatorname{Bdiag}\left(a_{11} \textbf{I}_{p_1}, \ldots, a_{KK} \textbf{I}_{p_K}\right)$, and $\bm{\Sigma}_{\bm{f}} = \left(b_{kk'}\right)$. 
	Due to the page limit, we present the results in the \href{Supplementary Material.pdf}{Supplementary Material}. 
	In cases where $p < n$, the proposed estimators demonstrate superior performance compared to the estimates produced by the R packages ``sem'', ``OpenMx'', and ``lavaan'' with lower average standard errors and much shorter computational time.
	When $p > 120$, the computational times of conventional methods become long and even intractable, resulting in ``NA'' values for model parameters.
	This is primarily due to the presence of singular sample covariance matrices.
	In general, the performance of all methods is comparable, with computational cost being the bottleneck for the competing methods.
	
	\begin{table}[!htb]
		\centering
		\caption{Estimation results ($\times 100$) using the proposed method for $a_{kk}$ and $b_{k k'}$ across $100$ replicates, given various $p$ and $n$, 
			where 
			``bias'' denotes the average of estimation bias, 
			``MCSD'' denotes the Monte Carlo standard deviation, 
			``ASE'' denotes the average standard error, 
			``$95\%$ CP'' denotes the coverage probability based on a $95\%$ Wald-type confidence interval.
		}
		\tiny
		\begin{tabular}{crrrrrrrrrrrrrr}
			\hline
			& \multicolumn{1}{c}{bias} & \multicolumn{1}{c}{MCSD} & \multicolumn{1}{c}{ASE} & \multicolumn{1}{c}{$95\%$ CP} & \multicolumn{1}{c}{} & \multicolumn{1}{c}{bias} & \multicolumn{1}{c}{MCSD} & \multicolumn{1}{c}{ASE} & \multicolumn{1}{c}{$95\%$ CP} & \multicolumn{1}{c}{} & \multicolumn{1}{c}{bias} & \multicolumn{1}{c}{MCSD} & \multicolumn{1}{c}{ASE} & \multicolumn{1}{c}{$95\%$ CP} \\ \cline{1-5} \cline{7-10} \cline{12-15} 
			$n = 40$ & \multicolumn{4}{c}{$2 \times (3,3,4) ^ \top$, $p = 20$} & \multicolumn{1}{c}{} & \multicolumn{4}{c}{$3 \times (3,3,4) ^ \top$, $p = 30$} & \multicolumn{1}{c}{} & \multicolumn{4}{c}{$4 \times (3,3,4) ^ \top$, $p = 40$} \\ \hline
			$a_{11}$ & 0.0 & 1.0 & 1.0 & 92 &  & 0.0 & 0.7 & 0.8 & 93 &  & 0.0 & 0.6 & 0.7 & 97 \\
			$a_{22}$ & -0.1 & 2.1 & 2.0 & 92 &  & 0.2 & 1.6 & 1.6 & 97 &  & -0.3 & 1.3 & 1.3 & 97 \\
			$a_{33}$ & -0.4 & 3.7 & 4.2 & 96 &  & 0.1 & 3.5 & 3.4 & 94 &  & -0.1 & 2.8 & 2.9 & 96 \\
			$b_{11}$ & -2.5 & 45.9 & 45.6 & 89 &  & -0.3 & 52.1 & 46.0 & 91 &  & 0.7 & 46.7 & 46.1 & 96 \\
			$b_{12}$ & -3.0 & 41.4 & 40.9 & 92 &  & -4.3 & 43.3 & 41.6 & 92 &  & -5.4 & 47.1 & 41.8 & 94 \\
			$b_{13}$ & 2.8 & 45.5 & 48.0 & 94 &  & 2.6 & 50.4 & 47.5 & 92 &  & -3.6 & 51.6 & 47.2 & 89 \\
			$b_{22}$ & -14.2 & 76.1 & 68.4 & 90 &  & -2.5 & 78.3 & 70.8 & 88 &  & -2.5 & 66.9 & 70.7 & 92 \\
			$b_{23}$ & -7.9 & 62.9 & 59.5 & 90 &  & -2.8 & 58.9 & 59.7 & 96 &  & -3.7 & 64.1 & 59.7 & 91 \\
			$b_{33}$ & 3.3 & 74.8 & 85.8 & 97 &  & -7.4 & 80.0 & 82.9 & 94 &  & -6.2 & 84.7 & 82.9 & 89 \\ \cline{1-5} \cline{7-10} \cline{12-15} 
			$n = 80$ & \multicolumn{4}{c}{$4 \times (3,3,4) ^ \top$, $p = 40$} & \multicolumn{1}{c}{} & \multicolumn{4}{c}{$8 \times (3,3,4) ^ \top$, $p = 80$} & \multicolumn{1}{c}{} & \multicolumn{4}{c}{$12 \times (3,3,4) ^ \top$, $p = 120$} \\ \hline
			$a_{11}$ & 0.0 & 0.5 & 0.5 & 95 &  & 0.1 & 0.4 & 0.3 & 93 &  & 0.0 & 0.2 & 0.3 & 98 \\
			$a_{22}$ & 0.0 & 0.9 & 1.0 & 97 &  & 0.0 & 0.7 & 0.7 & 93 &  & 0.0 & 0.5 & 0.5 & 96 \\
			$a_{33}$ & -0.4 & 1.9 & 2.0 & 95 &  & 0.0 & 1.5 & 1.4 & 96 &  & -0.1 & 1.3 & 1.2 & 93 \\
			$b_{11}$ & 3.1 & 33.3 & 32.8 & 92 &  & 1.0 & 28.8 & 32.4 & 97 &  & 0.4 & 31.6 & 32.3 & 95 \\
			$b_{12}$ & -0.1 & 31.3 & 29.6 & 95 &  & 0.3 & 28.4 & 29.7 & 97 &  & 1.9 & 26.6 & 29.6 & 96 \\
			$b_{13}$ & -0.4 & 35.5 & 33.2 & 96 &  & -0.2 & 33.2 & 33.2 & 96 &  & 2.0 & 33.4 & 33.8 & 95 \\
			$b_{22}$ & -5.3 & 48.3 & 49.2 & 94 &  & 3.3 & 50.1 & 50.5 & 95 &  & 3.0 & 54.0 & 50.4 & 92 \\
			$b_{23}$ & -3.9 & 41.0 & 41.5 & 93 &  & -3.6 & 44.1 & 42.2 & 93 &  & 6.4 & 42.8 & 43.2 & 97 \\
			$b_{33}$ & -13.2 & 55.1 & 57.2 & 93 &  & -7.4 & 56.1 & 57.8 & 92 &  & 6.3 & 60.4 & 59.9 & 96 \\ \cline{1-5} \cline{7-10} \cline{12-15} 
			$n = 120$ & \multicolumn{4}{c}{$4 \times (3,3,4) ^ \top$, $p = 40$} & \multicolumn{1}{c}{} & \multicolumn{4}{c}{$12 \times (3,3,4) ^ \top$, $p = 120$} & \multicolumn{1}{c}{} & \multicolumn{4}{c}{$20 \times (3,3,4) ^ \top$, $p = 200$} \\ \hline
			$a_{11}$ & -0.1 & 0.3 & 0.4 & 100 &  & 0.0 & 0.2 & 0.2 & 95 &  & 0.0 & 0.2 & 0.2 & 96 \\
			$a_{22}$ & -0.1 & 0.8 & 0.8 & 94 &  & -0.1 & 0.4 & 0.4 & 92 &  & 0.0 & 0.3 & 0.3 & 94 \\
			$a_{33}$ & 0.1 & 1.6 & 1.7 & 97 &  & -0.1 & 0.9 & 0.9 & 97 &  & 0.0 & 0.6 & 0.7 & 98 \\
			$b_{11}$ & -0.9 & 23.3 & 26.2 & 97 &  & -1.4 & 25.3 & 26.1 & 95 &  & -0.5 & 26.1 & 26.2 & 95 \\
			$b_{12}$ & 0.1 & 23.2 & 23.9 & 96 &  & -4.3 & 21.0 & 23.5 & 97 &  & 0.7 & 23.7 & 23.9 & 95 \\
			$b_{13}$ & -2.2 & 26.9 & 26.8 & 94 &  & -1.1 & 27.9 & 26.9 & 93 &  & -0.3 & 28.5 & 27.1 & 92 \\
			$b_{22}$ & -3.1 & 40.2 & 40.4 & 90 &  & -9.1 & 36.9 & 39.5 & 93 &  & -4.8 & 37.7 & 40.0 & 97 \\
			$b_{23}$ & -3.5 & 32.0 & 34.0 & 96 &  & -6.4 & 32.9 & 33.7 & 92 &  & -0.9 & 33.6 & 34.3 & 94 \\
			$b_{33}$ & -12.8 & 47.1 & 46.6 & 92 &  & -5.6 & 46.2 & 47.3 & 94 &  & -1.1 & 51.8 & 47.8 & 90 \\ \hline
		\end{tabular}
		\label{Tab:study_2}
	\end{table}
	
	\subsection{Simulation results with model misspecification} 
	
	\label{Subsec:study_3}
	
	We evaluate the performance of SCFA in scenarios of model misspecification, particularly when the covariance matrix deviates from an interconnected community structure.
	Under the misspecified structure, we generate $\bm{X}_i \sim N \left(\textbf{0}_{p \times 1}, \bm{\Sigma}_{\kappa} \right)$ with $\bm{\Sigma}_{\kappa} = \textbf{L} \bm{\Sigma}_{\bm{f}} \textbf{L} ^ \top + \bm{\Sigma}_{\bm{u}} + \textbf{E}_{\kappa}$, where $\textbf{L}$, $\bm{\Sigma}_{\bm{f}}$, and $\bm{\Sigma}_{\bm{u}}$ are model parameters defined in Section~\ref{Subsec:generation}.
	We set $n = 120$, $(p_1, \ldots, p_K) ^ \top = \left(60, 60, 80\right) ^ \top$, $K = 3$, and $p = 200$.
	$\textbf{E}_{\kappa}$ is an additional noise term that results in the covariance structure deviating from the interconnected community structure.
	Specifically, $\textbf{E}_{\kappa} \sim \text{Wishart}\left(p, \kappa \textbf{I}_p \right)$, where the noise level $\kappa \in 10 ^ {-2} \times \{1, 3, 5\}$ and the noise scale of each entry in $\textbf{E}_{\kappa}$ averages approximately $\{2, 6, 10\}$, which overwhelms $\bm{\Sigma}_{\bm{u}}$.
	We repeat this data generation procedure $100$ times.

	We apply the SCFA model to each simulated dataset and calculate the average bias, Monte Carlo standard deviation, average standard error, and $95\%$ coverage probability for each model parameter.
	The summarized results are presented in Table~\ref{Tab:study_3}.
	Additional results concerning higher values of $\kappa$ are available in the \href{Supplementary Material.pdf}{Supplementary Material}.
	
	The results demonstrate that the performance of the SCFA model remains robust with model misspecification. 
	In Table~\ref{Tab:study_3}, the estimates of covariance parameters among common factors $\widehat{\bm{\Sigma}}_{\bm{f}} = \left(\widehat{b}_{kk'}\right)$ are nearly invariant to the model misspecification. 
	In contrast, the estimates of error parameters $\widehat{a}_{kk}$ can reflect the noise by model misspecification, 
	and the bias is approximately the noise level. 
	Lastly, the estimated factors are also robust to the noise with less than $24\%$ relative Euclidean loss.

	\begin{table}[!htb]
		\centering
		\caption{Estimation results ($\times 100$) using the proposed method for $a_{kk}$ and $b_{kk'}$ under various noise terms $\textbf{E}_{\kappa}$, given $n = 120$, $(p_1, \ldots, p_K) ^ \top = (60, 60, 80) ^ \top$, $p = 200$,  
			where 
			``bias'' denotes the average of estimation bias, 
			``MCSD'' denotes the Monte Carlo standard deviation, 
			``ASE'' denotes the average standard error, 
			``$95\%$ CP'' denotes the coverage probability based on a $95\%$ Wald-type confidence interval.
			We also present the (average) relative Euclidean losses $\sum_{i = 1}^{n} \Vert \widehat{\bm{f}}_i - \bm{f}_i \Vert / \sum_{i = 1}^{n} \Vert \bm{f}_i \Vert \times 100\%$ for estimated factor scores across $100$ replicates under various noise levels. 
		}
		\tiny
		\begin{tabular}{crrrrrrrrrrrrrr}
			\hline
			& \multicolumn{4}{c}{noise scale $ = 2$} & \multicolumn{1}{c}{} & \multicolumn{4}{c}{noise scale $ = 6$} & \multicolumn{1}{c}{} & \multicolumn{4}{c}{noise scale $ = 10$} \\ \cline{2-5} \cline{7-10} \cline{12-15} 
			& \multicolumn{1}{c}{bias} & \multicolumn{1}{c}{MCSD} & \multicolumn{1}{c}{ASE} & \multicolumn{1}{c}{$95\%$ CP} & \multicolumn{1}{c}{} & \multicolumn{1}{c}{bias} & \multicolumn{1}{c}{MCSD} & \multicolumn{1}{c}{ASE} & \multicolumn{1}{c}{$95\%$ CP} & \multicolumn{1}{c}{} & \multicolumn{1}{c}{bias} & \multicolumn{1}{c}{MCSD} & \multicolumn{1}{c}{ASE} & \multicolumn{1}{c}{$95\%$ CP} \\ \hline
			$a_{11}$ & 203.2 & 3.8 & 3.6 & 0 &  & 602.3 & 13.4 & 10.3 & 0 &  & 989.9 & 21.4 & 16.9 & 0 \\
			$a_{22}$ & 203.5 & 4.6 & 3.8 & 0 &  & 596.1 & 13.2 & 10.4 & 0 &  & 986.7 & 19.5 & 17 & 0 \\
			$a_{33}$ & 201.9 & 4.5 & 3.7 & 0 &  & 614.1 & 11.8 & 9.7 & 0 &  & 1000.4 & 18.9 & 15.3 & 0 \\
			$b_{11}$ & -2.8 & 25.7 & 26.3 & 96 &  & -1.3 & 27 & 27.3 & 96 &  & -2.5 & 26.9 & 28 & 94 \\
			$b_{12}$ & 0.5 & 26.3 & 24.2 & 91 &  & 2.7 & 28.5 & 25.1 & 90 &  & 2.3 & 28.9 & 25.6 & 94 \\
			$b_{13}$ & -2.9 & 26.2 & 27.2 & 98 &  & -1.6 & 27.2 & 27.9 & 96 &  & 0.7 & 26.9 & 28.5 & 98 \\
			$b_{22}$ & 1.2 & 39.5 & 41.2 & 94 &  & 3.3 & 40.8 & 42.4 & 94 &  & 4.8 & 40.4 & 43.4 & 94 \\
			$b_{23}$ & -0.9 & 33.4 & 34.9 & 96 &  & 0.8 & 33.8 & 35.5 & 95 &  & 0.6 & 34 & 36.1 & 96 \\
			$b_{33}$ & 1.1 & 51.3 & 48.4 & 97 &  & 0.3 & 51.8 & 49 & 95 &  & 2.6 & 51.9 & 49.9 & 96 \\ \cline{1-5} \cline{7-10} \cline{12-15} 
			\multicolumn{1}{l}{relative loss} & 11\% & &  &  & \multicolumn{1}{l}{} & 18\% & &  &  & \multicolumn{1}{l}{} & 24\% & &  &  \\ \hline
		\end{tabular}
		\label{Tab:study_3}
	\end{table}
	
\section{Applications to Gene Expression Data}
	
\label{Sec:real}

	We apply SCFA to two RNA sequencing datasets from The Cancer Genome Atlas (TCGA) project \citep{KatarzynaPatrycjaMaciej2015} for pan-kidney cancer and glioblastoma multiforme (GBM) research. 
	High-throughput gene expression data often exhibit organized topological structures in co-expression networks. 
	We investigate whether our factor model can capture these latent network structures by representing correlated variables in a much-reduced number of factors.
	

\subsection{Pan-kidney cancer dataset}

\label{Subsec:real_1}

\begin{figure}[!htb]
		\centering
		\includegraphics[width = 0.7 \linewidth]{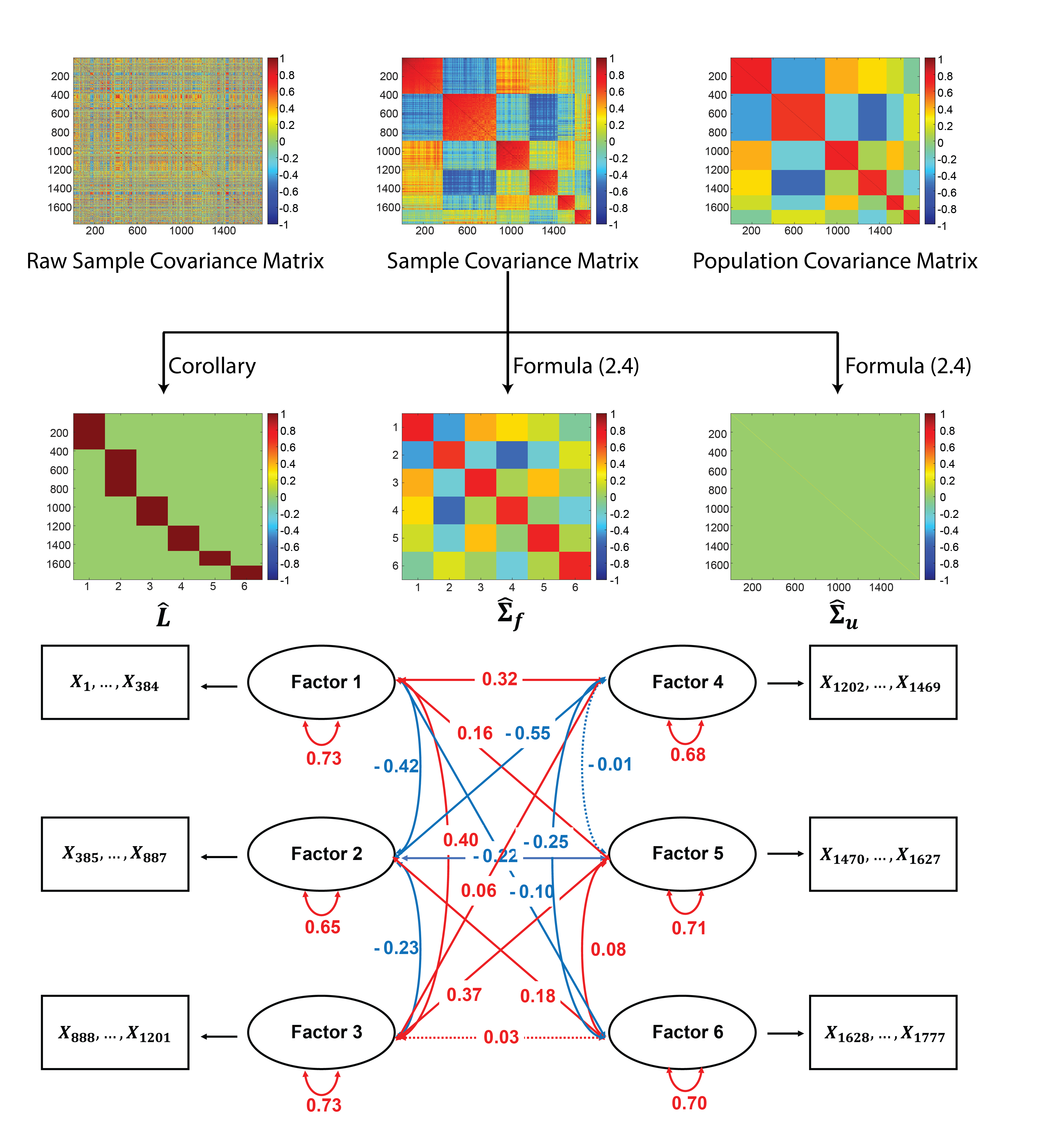} 
		\caption{
			The workflow (top) and the path diagram (bottom) for analyzing the TCGA kidney dataset.
			The subfigures in the first row of the workflow: 
			the input correlation matrix of $1777$ genes (left), 
			the reordered correlation matrix highlighting the detected interconnected community structure (middle), 
			and the (estimated) population correlation matrix (right),
			respectively. 
			In the second row: the estimates $\widehat{\textbf{L}}$ (left), $\widehat{\bm{\Sigma}}_{\bm{f}}$ (middle), and $\widehat{\bm{\Sigma}}_{\bm{u}}$ (right), respectively. 
			The bottom path diagram illustrates gene subsets (in rectangles) and their respective
			common factors (in oval circles): 
			the lines (with double-ended arrows) connecting the $6$ common factors to themselves denote their estimated variances (i.e., the diagonal entries of $\widehat{\bm{\Sigma}}_{\bm{f}}$);
			and the lines (with double-ended arrows) among the $6$ common factors denote their
			estimated covariances (i.e., the off-diagonal entries of $\widehat{\bm{\Sigma}}_{\bm{f}}$). 
			Specifically, the red solid lines represent
			significant positive estimates, the blue solid lines represent significant negative estimates, and the dashed lines represent non-significant estimates.
			%
		}
		\label{Fig:real_1}
	\end{figure}

	This gene expression dataset comprised information on the expression levels in Reads Per Kilobase of transcript per Million mapped reads (RPKM) for $13408$ genes, collected from $712$ kidney cancer patients with a mean age of $59.66$ years (standard deviation $12.59$ years) and a sex distribution of $34.27\%$ females and $65.73\%$ males.
	We performed data preprocessing using a standard pipeline. Specifically, genes with low average expression values ($\leq 5$) were removed from the matched sample. 
	In SCFA, we first extracted an interconnected community structure from the gene co-expression matrix using the algorithm by \citet{WuMaLiu2021}, while other community detection algorithms provided similar findings.  
	The detected interconnected community structure comprised $K = 6$ communities from $p = 1777$ genes, i.e., $\bm{X}_i$, with varying sizes: $p_1 = 384$, $p_2 = 503$, $p_3 = 314$, $p_4 = 268$, $p_5 = 158$, and $p_6 = 150$, as illustrated in Figure~\ref{Fig:real_1}. 
	While gene expression variables were highly correlated within each module, the modules themselves were also positively or negatively intercorrelated, as demonstrated in the top subfigures of Figure~\ref{Fig:real_1}. 
	Our goal was to estimate latent factors for these interconnected modules (i.e., dimension reduction of correlated genes) using SCFA.

	Based on the learned interconnected community structure, we applied our approach for parameter estimation. 
	The results were presented in Figure~\ref{Fig:real_1}. 
	The first row of Figure~\ref{Fig:real_1} (left to right) demonstrated the extracted interconnected community structure (middle) from the original data (left), and the large correlation matrix estimate (right) by \citet{YangChenChen2024}. 
	In the second row of Figure~\ref{Fig:real_1}, we showed the results of decomposing the above-estimated correlation matrix of $\bm{\Sigma}$ into SCFA parameters $\textbf{L} \bm{\Sigma}_{\bm{f}} \textbf{L} ^ \top + \bm{\Sigma}_{\bm{u}}$.
	We exhibited the estimated matrices $\widehat{\textbf{L}}$, $\widehat{\bm{\Sigma}}_{\bm{f}}$, and $\widehat{\bm{\Sigma}}_{\bm{u}}$ using heatmaps. 
	The middle heatmap in the second row illustrated the interactive relationships between $6$ common factors.
	The bottom subfigure of Figure~\ref{Fig:real_1} was a classic path diagram that demonstrates the result of factor analysis, including factor memberships, intra-factor, and inter-factor correlations. 
	As shown, SCFA reduced $p = 1777$ gene expression variables to $K = 6$ correlated common factors (with the estimated covariance matrix $\widehat{\bm{\Sigma}}_{\bm{f}}$), while accurately representing the complex relationships among all $1777$ variables. The computational time was less than one minute.
	Due to the page limit, we provided detailed numerical estimates, gene names, the corresponding community membership, and estimated factor scores in the \href{Supplementary Material.pdf}{Supplementary Material}.

	We further conducted pathway analysis to investigate the relationships between each common factor and kidney cancer. 
	The results revealed that each common factor characterized unique cellular and molecular functions related to kidney cancer.
	Specifically, the first common factor exhibited enrichment in pathways related to G protein-coupled receptors (GPCRs), that play pivotal roles in various aspects of cancer progression, including tumor growth, invasion, migration, survival, and metastasis \citep{ArakakiPanTrejo2018}.
	The second common factor was enriched with pathways related to cellular respiration in mitochondria, which are essential to cancer cells, including renal adenocarcinoma \citep{Wallace2012}. 
	The third common factor was enriched with pathways associated with the body's immune response: renal cell carcinoma (RCC) is considered to be an immunogenic tumor but is known to mediate immune dysfunction \citep{DiazMonteroRiniFinke2020}.
	For the remaining common factors, the uniqueness of associations might be less pronounced, and detailed information was provided in the \href{Supplementary Material.pdf}{Supplementary Material}.
	
	\subsection{GBM dataset}
	
	\label{Subsec:real_2}
	
\begin{figure}[!htb]
		\begin{center}
			\includegraphics[width = 0.7 \linewidth]{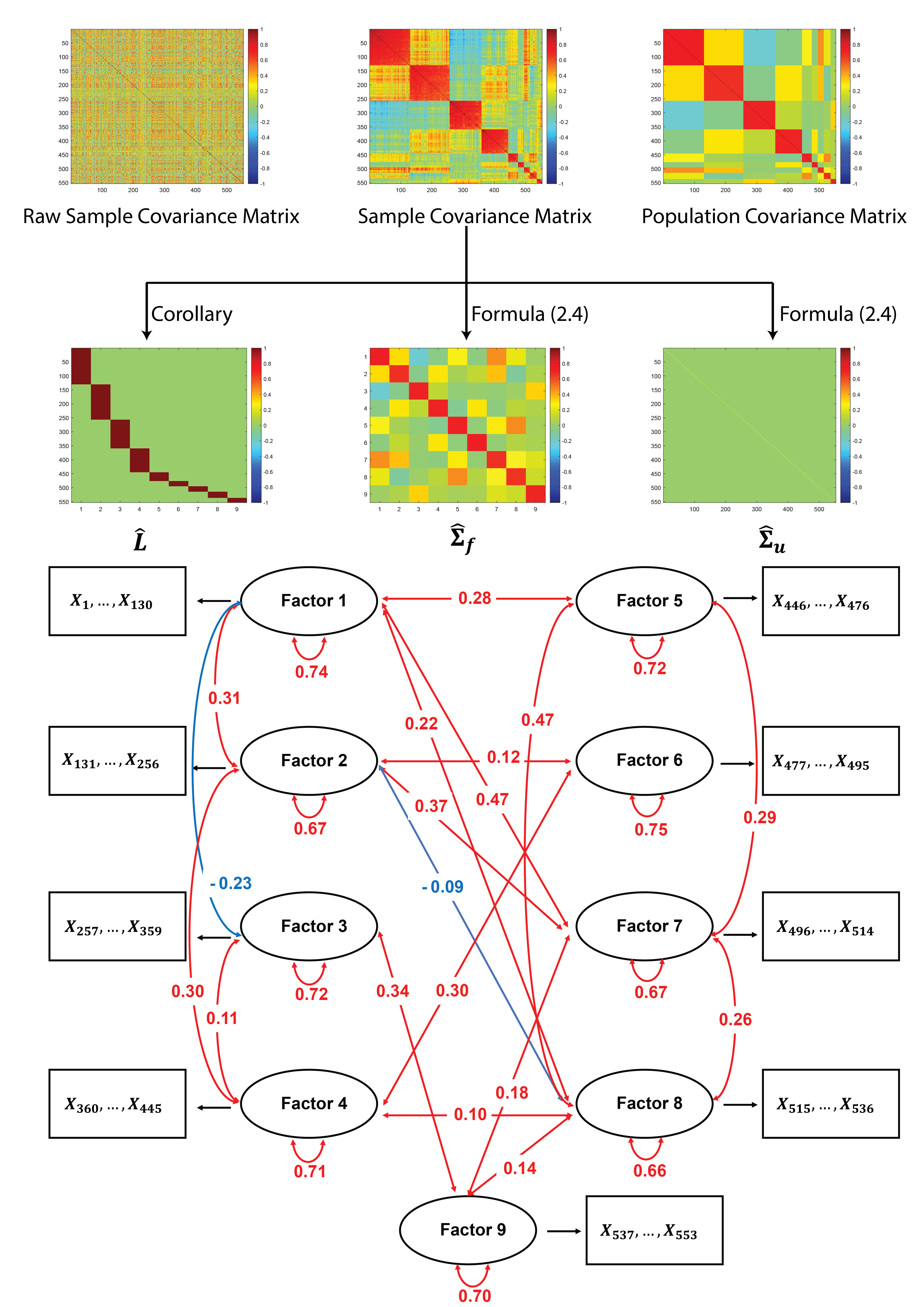} 
			\caption{
				The workflow (top) and the path diagram (bottom) for analyzing the TCGA GBM dataset.
				The subfigures in the first row of the workflow: 
				the input correlation matrix of $553$ genes (left), 
				the reordered correlation matrix highlighting the detected interconnected community structure (middle), 
				and the (estimated) population correlation matrix (right),
				respectively. 
				In the second row: the estimates $\widehat{\textbf{L}}$ (left), $\widehat{\bm{\Sigma}}_{\bm{f}}$ (middle), and $\widehat{\bm{\Sigma}}_{\bm{u}}$ (right), respectively. 
				The bottom path diagram illustrates gene subsets (in rectangles) and their respective
				common factors (in oval circles): 
				the lines (with double-ended arrows) connecting the $9$ common factors to themselves denote their estimated variances (i.e., the diagonal entries of $\widehat{\bm{\Sigma}}_{\bm{f}}$);
				and the lines (with double-ended arrows) among the $9$ common factors denote their
				estimated covariances (i.e., the off-diagonal entries of $\widehat{\bm{\Sigma}}_{\bm{f}}$). 
				Specifically, the red solid lines represent
				significant positive estimates and the blue solid lines represent significant negative estimates (the non-significant estimates are omitted, please see a complete version in the \href{Supplementary Material.pdf}{Supplementary Material}).
			}
			\label{Fig:real_2}
		\end{center}
	\end{figure}

	The second dataset measured the RNA-seq data on $8196$ genes from $n = 278$ GBM patients, with a mean age of $57.73$ years (standard deviation $15.02$ years) and a sex distribution of $40\%$ females and $60\%$ males.
	The data processing was performed similarly to the first dataset.
	We calculated an interconnected community structure using the interconnected community detection algorithm and observed $K = 9$ communities formed by $p = 553$ genes, i.e., $\bm{X}_i$, with sizes of $p_1 = 130$, $p_2 = 126$, $p_3 = 103$, $p_4 = 86$, $p_5 = 31$, $p_6 = 19$, $p_7 = 19$, $p_8 = 22$, and $p_9 = 17$.
	We presented the results using the heatmaps in Figure~\ref{Fig:real_2}.
	Similarly, our goal was to explore the latent common factors and conduct statistical inference about the relationships between these common factors.

	Using the extracted interconnected community structure, we adopted the SCFA model to TCGA GBM dataset and presented the estimation results in Figure~\ref{Fig:real_2}. 
	The subfigures in the first row of the workflow demonstrated the original sample correlation matrix (left), the learned interconnected community structure (middle), and the estimated correlation matrix (right) using the algorithm \citep{YangChenChen2024}, respectively.
	Those in the second row of the workflow showed the decomposition results, i.e., $\widehat{\textbf{L}}$, $\widehat{\bm{\Sigma}}_{\bm{f}}$, and $\widehat{\bm{\Sigma}}_{\bm{u}}$, using the heatmaps. 
	In particular, the heatmap of $\widehat{\bm{\Sigma}}_{\bm{f}}$ illustrated the interactive relationships between $9$ common factors.
	The bottom path diagram in Figure~\ref{Fig:real_2} demonstrated the factor analysis results.
	We provided more results in the \href{Supplementary Material.pdf}{Supplementary Material}.
		
	We primarily focused on the pathway analysis for the first four and the sixth common factors, which had large numbers of genes. 
	Specifically, most genes in the first common factor were related to cell cycle, which is reasonable for most cancers, including TCGA GBM \citep{BrennanVerhaakMcKenna2013}.
	The second common factor was broader, encompassing RNA transcription and metabolic processes. 
	The third one exhibited a relationship with the immune system, which is relevant to most cancers. 
	The fourth common factor was enriched with pathways related to cell adhesion, cell-to-cell signaling, nervous system, and synapse, which, although general to many cancer types, might be more specific to GBM.
	Many genes in the sixth common factors were related to neurotransmitter and neurological process, which were very likely to be specific to GBM.

	For comparison, we also applied the conventional CFA model with the same detected community membership to both datasets using R packages ``sem”, ``OpenMx”, and ``lavaan”, respectively. 
	However, none of these standard computational packages could handle these gene expression datasets due to $n = 712 < p = 1777$ and $n = 278 < p = 553$.
	Thus, the SCFA model fills the methodological gap in performing CFA for high-dimensional datasets.
	
\section{Discussion}
	
\label{Sec:discussion}
	
	We have developed a novel confirmatory factor analysis model, the SCFA model, designed for high-dimensional data.
	In contrast to the traditional CFA model equipped with standard computational packages, which relies on predefined model specifications and faces computational challenges with high-dimensional datasets, the SCFA model simultaneously addresses these limitations by leveraging a data-driven covariance structure. 
	Due to its prevalence and block structure, we integrate the ubiquitous interconnected community structure observed in covariance matrices across diverse high-dimensional data types into the SCFA model.
	This integration yields likelihood-based UMVUEs for the factor loading matrix $\textbf{L}$, the covariance matrix of common factors $\bm{\Sigma}_{\bm{f}}$, and the covariance matrix of error terms $\bm{\Sigma}_{\bm{u}}$ in closed forms, as well as explicitly consistent least-square estimators for the factor scores. 
	Additionally, explicit variance estimators further facilitate statistical inference for these parameters. 
	
	Extensive simulation studies have demonstrated the superior performance of the SCFA model in both parameter estimation and factor score accuracy.
	Notably, SCFA significantly reduces the computational burden compared to existing CFA computational packages. 
	The SCFA model also exhibits robustness to moderate violations of the interconnected community structure. 
	In applications to two TCGA gene expression datasets, we employed the SCFA model to explore both relationships between $6$ common factors and $1777$ genes in the kidney data, and between $9$ common factors and $553$ genes in the GBM data, conducting statistical inference on covariances between these common factors.
	In both datasets, highly correlated gene expression variables form inter-connected communities can be represented by common factors and the covariance matrix of common factors $\bm{\Sigma}_{\bm{f}}$. 
	Therefore, the confirmatory factor model is a natural fit for the co-expression patterns in omics data and SCFA provides a computationally efficient and accurate solution.

	In summary, SCFA models combine the benefits of conventional CFA models, including dimension reduction and flexibility in covariance matrix modeling for common factors, with the ability to adapt to data-driven covariance structures. 
	As many high-dimensional datasets implicitly exhibit interconnected community structures, SCFA holds broad applicability.
	
\begin{acks}[Acknowledgments]
The authors would like to thank the anonymous referees, an Associate Editor and the Editor for their constructive comments that improved the quality of this paper.
\end{acks}
%

\begin{supplement}
\stitle{Supplementary Material (pdf file)}
\sdescription{Supplementary material contains the definitions and properties of the uniform-block matrices, the additional numerical results, the additional simulation studies, and all technical proofs.}
\end{supplement}
\begin{supplement}
\stitle{SCFA Code (zipped file)}
\sdescription{The folder contains all R code for the figures, simulation studies, and real data analysis.}
\end{supplement}


\bibliographystyle{imsart-nameyear} 
\bibliography{paper_ref_SCFA}       

\begin{thebibliography}{49}

\bibitem[\protect\citeauthoryear{Anderson}{2003}]{Anderson2003}
\begin{bbook}[author]
\bauthor{\bsnm{Anderson},~\bfnm{T.~W.}\binits{T.~W.}}
(\byear{2003}).
\btitle{An introduction to multivariate statistical analysis},
\bedition{third} ed.
\bseries{Wiley Series in Probability and Statistics}.
\bpublisher{Wiley-Interscience [John Wiley \& Sons], Hoboken, NJ}.
\bmrnumber{1990662}
\end{bbook}
\endbibitem

\bibitem[\protect\citeauthoryear{Arakaki, Pan and
  Trejo}{2018}]{ArakakiPanTrejo2018}
\begin{barticle}[author]
\bauthor{\bsnm{Arakaki},~\bfnm{Aleena K.~S.}\binits{A.~K.~S.}},
  \bauthor{\bsnm{Pan},~\bfnm{Wen-An}\binits{W.-A.}} \AND
  \bauthor{\bsnm{Trejo},~\bfnm{JoAnn}\binits{J.}}
(\byear{2018}).
\btitle{GPCRs in Cancer: Protease-Activated Receptors, Endocytic Adaptors and
  Signaling}.
\bjournal{International Journal of Molecular Sciences}
\bvolume{19}.
\bdoi{10.3390/ijms19071886}
\end{barticle}
\endbibitem

\bibitem[\protect\citeauthoryear{Bai and Li}{2012}]{BaiLi2012}
\begin{barticle}[author]
\bauthor{\bsnm{Bai},~\bfnm{Jushan}\binits{J.}} \AND
  \bauthor{\bsnm{Li},~\bfnm{Kunpeng}\binits{K.}}
(\byear{2012}).
\btitle{Statistical analysis of factor models of high dimension}.
\bjournal{Ann. Statist.}
\bvolume{40}
\bpages{436--465}.
\bdoi{10.1214/11-AOS966}
\bmrnumber{3014313}
\end{barticle}
\endbibitem

\bibitem[\protect\citeauthoryear{Basilevsky}{2009}]{Basilevsky2009}
\begin{bbook}[author]
\bauthor{\bsnm{Basilevsky},~\bfnm{Alexander}\binits{A.}}
(\byear{2009}).
\btitle{Statistical factor analysis and related methods}.
\bseries{Wiley Series in Probability and Mathematical Statistics: Probability
  and Mathematical Statistics}.
\bpublisher{John Wiley \& Sons, Inc., New York}
\bnote{Theory and applications, A Wiley-Interscience Publication}.
\bdoi{10.1002/9780470316894}
\bmrnumber{1277625}
\end{bbook}
\endbibitem

\bibitem[\protect\citeauthoryear{Boker et~al.}{2023}]{RBokerNealeMaes2023}
\begin{bmanual}[author]
\bauthor{\bsnm{Boker},~\bfnm{Steven~M.}\binits{S.~M.}},
  \bauthor{\bsnm{Neale},~\bfnm{Michael~C.}\binits{M.~C.}},
  \bauthor{\bsnm{Maes},~\bfnm{Hermine~H.}\binits{H.~H.}},
  \bauthor{\bsnm{Wilde},~\bfnm{Michael~J.}\binits{M.~J.}},
  \bauthor{\bsnm{Spiegel},~\bfnm{Michael}\binits{M.}},
  \bauthor{\bsnm{Brick},~\bfnm{Timothy~R.}\binits{T.~R.}},
  \bauthor{\bsnm{Estabrook},~\bfnm{Ryne}\binits{R.}},
  \bauthor{\bsnm{Bates},~\bfnm{Timothy~C.}\binits{T.~C.}},
  \bauthor{\bsnm{Mehta},~\bfnm{Paras}\binits{P.}}, \bauthor{\bparticle{von}
  \bsnm{Oertzen},~\bfnm{Timo}\binits{T.}},
  \bauthor{\bsnm{Gore},~\bfnm{Ross~J.}\binits{R.~J.}},
  \bauthor{\bsnm{Hunter},~\bfnm{Michael~D.}\binits{M.~D.}},
  \bauthor{\bsnm{Hackett},~\bfnm{Daniel~C.}\binits{D.~C.}},
  \bauthor{\bsnm{Karch},~\bfnm{Julian}\binits{J.}},
  \bauthor{\bsnm{Brandmaier},~\bfnm{Andreas~M.}\binits{A.~M.}},
  \bauthor{\bsnm{Pritikin},~\bfnm{Joshua~N.}\binits{J.~N.}},
  \bauthor{\bsnm{Zahery},~\bfnm{Mahsa}\binits{M.}},
  \bauthor{\bsnm{Kirkpatrick},~\bfnm{Robert~M.}\binits{R.~M.}},
  \bauthor{\bsnm{Wang},~\bfnm{Yang}\binits{Y.}},
  \bauthor{\bsnm{Goodrich},~\bfnm{Ben}\binits{B.}},
  \bauthor{\bsnm{Driver},~\bfnm{Charles}\binits{C.}}, \bauthor{\bparticle{of}
  \bsnm{Technology},~\bfnm{Massachusetts~Institute}\binits{M.~I.}},
  \bauthor{\bsnm{Johnson},~\bfnm{S.~G.}\binits{S.~G.}},
  \bauthor{\bparticle{for}
  \bsnm{Computing~Machinery},~\bfnm{Association}\binits{A.}},
  \bauthor{\bsnm{Kraft},~\bfnm{Dieter}\binits{D.}},
  \bauthor{\bsnm{Wilhelm},~\bfnm{Stefan}\binits{S.}},
  \bauthor{\bsnm{Medland},~\bfnm{Sarah}\binits{S.}},
  \bauthor{\bsnm{Falk},~\bfnm{Carl~F.}\binits{C.~F.}},
  \bauthor{\bsnm{Keller},~\bfnm{Matt}\binits{M.}},
  \bauthor{\bsnm{G},~\bfnm{Manjunath~B}\binits{M.~B.}},
  \bauthor{\bparticle{of~the University~of}
  \bsnm{California},~\bfnm{The~Regents}\binits{T.~R.}},
  \bauthor{\bsnm{Ingber},~\bfnm{Lester}\binits{L.}},
  \bauthor{\bsnm{Voon},~\bfnm{Wong~Shao}\binits{W.~S.}},
  \bauthor{\bsnm{Palacios},~\bfnm{Juan}\binits{J.}},
  \bauthor{\bsnm{Yang},~\bfnm{Jiang}\binits{J.}},
  \bauthor{\bsnm{Guennebaud},~\bfnm{Gael}\binits{G.}} \AND
  \bauthor{\bsnm{Niesen},~\bfnm{Jitse}\binits{J.}}
(\byear{2023}).
\btitle{Extended Structural Equation Modelling}
\bnote{R package version 2.21.8}.
\end{bmanual}
\endbibitem

\bibitem[\protect\citeauthoryear{Brennan
  et~al.}{2013}]{BrennanVerhaakMcKenna2013}
\begin{barticle}[author]
\bauthor{\bsnm{Brennan},~\bfnm{Cameron~W}\binits{C.~W.}},
  \bauthor{\bsnm{Verhaak},~\bfnm{Roel~GW}\binits{R.~G.}},
  \bauthor{\bsnm{McKenna},~\bfnm{Aaron}\binits{A.}},
  \bauthor{\bsnm{Campos},~\bfnm{Benito}\binits{B.}},
  \bauthor{\bsnm{Noushmehr},~\bfnm{Houtan}\binits{H.}},
  \bauthor{\bsnm{Salama},~\bfnm{Sofie~R}\binits{S.~R.}},
  \bauthor{\bsnm{Zheng},~\bfnm{Siyuan}\binits{S.}},
  \bauthor{\bsnm{Chakravarty},~\bfnm{Debyani}\binits{D.}},
  \bauthor{\bsnm{Sanborn},~\bfnm{J~Zachary}\binits{J.~Z.}},
  \bauthor{\bsnm{Berman},~\bfnm{Samuel~H}\binits{S.~H.}} \betal{et~al.}
(\byear{2013}).
\btitle{The somatic genomic landscape of glioblastoma}.
\bjournal{cell}
\bvolume{155}
\bpages{462--477}.
\end{barticle}
\endbibitem

\bibitem[\protect\citeauthoryear{Brown}{2015}]{Brown2015}
\begin{bbook}[author]
\bauthor{\bsnm{Brown},~\bfnm{Timothy~A}\binits{T.~A.}}
(\byear{2015}).
\btitle{Confirmatory factor analysis for applied research}.
\bpublisher{Guilford publications}.
\end{bbook}
\endbibitem

\bibitem[\protect\citeauthoryear{Browne}{2001}]{Browne2001}
\begin{barticle}[author]
\bauthor{\bsnm{Browne},~\bfnm{Michael~W.}\binits{M.~W.}}
(\byear{2001}).
\btitle{An Overview of Analytic Rotation in Exploratory Factor Analysis}.
\bjournal{Multivariate Behavioral Research}
\bvolume{36}
\bpages{111--150}.
\bdoi{10.1207/S15327906MBR3601\_05}
\end{barticle}
\endbibitem

\bibitem[\protect\citeauthoryear{Carlson and Mulaik}{1993}]{CarlsonMulaik1993}
\begin{barticle}[author]
\bauthor{\bsnm{Carlson},~\bfnm{Marianne}\binits{M.}} \AND
  \bauthor{\bsnm{Mulaik},~\bfnm{Stanley~A}\binits{S.~A.}}
(\byear{1993}).
\btitle{Trait ratings from descriptions of behavior as mediated by components
  of meaning}.
\bjournal{Multivariate Behavioral Research}
\bvolume{28}
\bpages{111--159}.
\end{barticle}
\endbibitem

\bibitem[\protect\citeauthoryear{Chen, Fan and Zhu}{2023}]{ChenFanZhu2023}
\begin{barticle}[author]
\bauthor{\bsnm{Chen},~\bfnm{Elynn~Y.}\binits{E.~Y.}},
  \bauthor{\bsnm{Fan},~\bfnm{Jianqing}\binits{J.}} \AND
  \bauthor{\bsnm{Zhu},~\bfnm{Xuening}\binits{X.}}
(\byear{2023}).
\btitle{Community network auto-regression for high-dimensional time series}.
\bjournal{J. Econometrics}
\bvolume{235}
\bpages{1239--1256}.
\bdoi{10.1016/j.jeconom.2022.10.005}
\bmrnumber{4602909}
\end{barticle}
\endbibitem

\bibitem[\protect\citeauthoryear{Chen et~al.}{2018}]{ChenKangXing2018}
\begin{barticle}[author]
\bauthor{\bsnm{Chen},~\bfnm{Shuo}\binits{S.}},
  \bauthor{\bsnm{Kang},~\bfnm{Jian}\binits{J.}},
  \bauthor{\bsnm{Xing},~\bfnm{Yishi}\binits{Y.}},
  \bauthor{\bsnm{Zhao},~\bfnm{Yunpeng}\binits{Y.}} \AND
  \bauthor{\bsnm{Milton},~\bfnm{Donald~K.}\binits{D.~K.}}
(\byear{2018}).
\btitle{Estimating large covariance matrix with network topology for
  high-dimensional biomedical data}.
\bjournal{Comput. Statist. Data Anal.}
\bvolume{127}
\bpages{82--95}.
\bdoi{10.1016/j.csda.2018.05.008}
\bmrnumber{3820311}
\end{barticle}
\endbibitem

\bibitem[\protect\citeauthoryear{Chen et~al.}{2024}]{ChenZhangWu2024}
\begin{barticle}[author]
\bauthor{\bsnm{Chen},~\bfnm{Shuo}\binits{S.}},
  \bauthor{\bsnm{Zhang},~\bfnm{Yuan}\binits{Y.}},
  \bauthor{\bsnm{Wu},~\bfnm{Qiong}\binits{Q.}},
  \bauthor{\bsnm{Bi},~\bfnm{Chuan}\binits{C.}},
  \bauthor{\bsnm{Kochunov},~\bfnm{Peter}\binits{P.}} \AND
  \bauthor{\bsnm{Hong},~\bfnm{L~Elliot}\binits{L.~E.}}
(\byear{2024}).
\btitle{Identifying covariate-related subnetworks for whole-brain connectome
  analysis}.
\bjournal{Biostatistics}
\bvolume{25}
\bpages{541--558}.
\bdoi{10.1093/biostatistics/kxad007}
\bmrnumber{4732247}
\end{barticle}
\endbibitem

\bibitem[\protect\citeauthoryear{Chiappelli
  et~al.}{2019}]{ChiappelliRowlandWijtenburg2019}
\begin{barticle}[author]
\bauthor{\bsnm{Chiappelli},~\bfnm{Joshua}\binits{J.}},
  \bauthor{\bsnm{Rowland},~\bfnm{Laura~M.}\binits{L.~M.}},
  \bauthor{\bsnm{Wijtenburg},~\bfnm{S.~Andrea}\binits{S.~A.}},
  \bauthor{\bsnm{Chen},~\bfnm{Hongji}\binits{H.}},
  \bauthor{\bsnm{Maudsley},~\bfnm{Andrew~A.}\binits{A.~A.}},
  \bauthor{\bsnm{Sheriff},~\bfnm{Sulaiman}\binits{S.}},
  \bauthor{\bsnm{Chen},~\bfnm{Shuo}\binits{S.}},
  \bauthor{\bsnm{Savransky},~\bfnm{Anya}\binits{A.}},
  \bauthor{\bsnm{Marshall},~\bfnm{Wyatt}\binits{W.}},
  \bauthor{\bsnm{Ryan},~\bfnm{Meghann~C.}\binits{M.~C.}},
  \bauthor{\bsnm{Bruce},~\bfnm{Heather~A.}\binits{H.~A.}},
  \bauthor{\bsnm{Shuldiner},~\bfnm{Alan~R.}\binits{A.~R.}},
  \bauthor{\bsnm{Mitchell},~\bfnm{Braxton~D.}\binits{B.~D.}},
  \bauthor{\bsnm{Kochunov},~\bfnm{Peter}\binits{P.}} \AND
  \bauthor{\bsnm{Hong},~\bfnm{L.~Elliot}\binits{L.~E.}}
(\byear{2019}).
\btitle{Cardiovascular risks impact human brain N-acetylaspartate in regionally
  specific patterns}.
\bjournal{Proceedings of the National Academy of Sciences}
\bvolume{116}
\bpages{25243--25249}.
\bdoi{10.1073/pnas.1907730116}
\end{barticle}
\endbibitem

\bibitem[\protect\citeauthoryear{Colizza
  et~al.}{2006}]{ColizzaFlamminiSerrano2006}
\begin{barticle}[author]
\bauthor{\bsnm{Colizza},~\bfnm{Vittoria}\binits{V.}},
  \bauthor{\bsnm{Flammini},~\bfnm{Alessandro}\binits{A.}},
  \bauthor{\bsnm{Serrano},~\bfnm{M~Angeles}\binits{M.~A.}} \AND
  \bauthor{\bsnm{Vespignani},~\bfnm{Alessandro}\binits{A.}}
(\byear{2006}).
\btitle{Detecting rich-club ordering in complex networks}.
\bjournal{Nature physics}
\bvolume{2}
\bpages{110--115}.
\end{barticle}
\endbibitem

\bibitem[\protect\citeauthoryear{D{\'\i}az-Montero, Rini and
  Finke}{2020}]{DiazMonteroRiniFinke2020}
\begin{barticle}[author]
\bauthor{\bsnm{D{\'\i}az-Montero},~\bfnm{C~Marcela}\binits{C.~M.}},
  \bauthor{\bsnm{Rini},~\bfnm{Brian~I}\binits{B.~I.}} \AND
  \bauthor{\bsnm{Finke},~\bfnm{James~H}\binits{J.~H.}}
(\byear{2020}).
\btitle{The immunology of renal cell carcinoma}.
\bjournal{Nature Reviews Nephrology}
\bvolume{16}
\bpages{721--735}.
\end{barticle}
\endbibitem

\bibitem[\protect\citeauthoryear{Fan and Han}{2017}]{FanHan2017}
\begin{barticle}[author]
\bauthor{\bsnm{Fan},~\bfnm{Jianqing}\binits{J.}} \AND
  \bauthor{\bsnm{Han},~\bfnm{Xu}\binits{X.}}
(\byear{2017}).
\btitle{Estimation of the false discovery proportion with unknown dependence}.
\bjournal{J. R. Stat. Soc. Ser. B. Stat. Methodol.}
\bvolume{79}
\bpages{1143--1164}.
\bdoi{10.1111/rssb.12204}
\bmrnumber{3689312}
\end{barticle}
\endbibitem

\bibitem[\protect\citeauthoryear{Fan, Liao and
  Mincheva}{2013}]{FanLiaoMincheva2013}
\begin{barticle}[author]
\bauthor{\bsnm{Fan},~\bfnm{Jianqing}\binits{J.}},
  \bauthor{\bsnm{Liao},~\bfnm{Yuan}\binits{Y.}} \AND
  \bauthor{\bsnm{Mincheva},~\bfnm{Martina}\binits{M.}}
(\byear{2013}).
\btitle{Large covariance estimation by thresholding principal orthogonal
  complements}.
\bjournal{J. R. Stat. Soc. Ser. B. Stat. Methodol.}
\bvolume{75}
\bpages{603--680}.
\bnote{With 33 discussions by 57 authors and a reply by Fan, Liao and
  Mincheva}.
\bdoi{10.1111/rssb.12016}
\bmrnumber{3091653}
\end{barticle}
\endbibitem

\bibitem[\protect\citeauthoryear{Fan et~al.}{2020}]{FanLiZhang2020}
\begin{bbook}[author]
\bauthor{\bsnm{Fan},~\bfnm{Jianqing}\binits{J.}},
  \bauthor{\bsnm{Li},~\bfnm{Runze}\binits{R.}},
  \bauthor{\bsnm{Zhang},~\bfnm{Cun-Hui}\binits{C.-H.}} \AND
  \bauthor{\bsnm{Zou},~\bfnm{Hui}\binits{H.}}
(\byear{2020}).
\btitle{Statistical foundations of data science}.
\bpublisher{Chapman and Hall/CRC}.
\end{bbook}
\endbibitem

\bibitem[\protect\citeauthoryear{Fan et~al.}{2021}]{FanWangZhong2021}
\begin{barticle}[author]
\bauthor{\bsnm{Fan},~\bfnm{Jianqing}\binits{J.}},
  \bauthor{\bsnm{Wang},~\bfnm{Kaizheng}\binits{K.}},
  \bauthor{\bsnm{Zhong},~\bfnm{Yiqiao}\binits{Y.}} \AND
  \bauthor{\bsnm{Zhu},~\bfnm{Ziwei}\binits{Z.}}
(\byear{2021}).
\btitle{Robust high-dimensional factor models with applications to statistical
  machine learning}.
\bjournal{Statist. Sci.}
\bvolume{36}
\bpages{303--327}.
\bdoi{10.1214/20-sts785}
\bmrnumber{4255196}
\end{barticle}
\endbibitem

\bibitem[\protect\citeauthoryear{Fortunato}{2010}]{Fortunato2010}
\begin{barticle}[author]
\bauthor{\bsnm{Fortunato},~\bfnm{Santo}\binits{S.}}
(\byear{2010}).
\btitle{Community detection in graphs}.
\bjournal{Phys. Rep.}
\bvolume{486}
\bpages{75--174}.
\bdoi{10.1016/j.physrep.2009.11.002}
\bmrnumber{2580414}
\end{barticle}
\endbibitem

\bibitem[\protect\citeauthoryear{Fox}{2006}]{Fox2006}
\begin{barticle}[author]
\bauthor{\bsnm{Fox},~\bfnm{John}\binits{J.}}
(\byear{2006}).
\btitle{Structural equation modeling with the sem package in {R}}.
\bjournal{Struct. Equ. Model.}
\bvolume{13}
\bpages{465--486}.
\bdoi{10.1207/s15328007sem1303\_7}
\bmrnumber{2240110}
\end{barticle}
\endbibitem

\bibitem[\protect\citeauthoryear{Fox et~al.}{2022}]{RFoxNieByrnes2022}
\begin{bmanual}[author]
\bauthor{\bsnm{Fox},~\bfnm{John}\binits{J.}},
  \bauthor{\bsnm{Nie},~\bfnm{Zhenghua}\binits{Z.}},
  \bauthor{\bsnm{Byrnes},~\bfnm{Jarrett}\binits{J.}},
  \bauthor{\bsnm{Culbertson},~\bfnm{Michael}\binits{M.}},
  \bauthor{\bsnm{DebRoy},~\bfnm{Saikat}\binits{S.}},
  \bauthor{\bsnm{Friendly},~\bfnm{Michael}\binits{M.}},
  \bauthor{\bsnm{Goodrich},~\bfnm{Benjamin}\binits{B.}},
  \bauthor{\bsnm{Jones},~\bfnm{Richard~H.}\binits{R.~H.}},
  \bauthor{\bsnm{Kramer},~\bfnm{Adam}\binits{A.}},
  \bauthor{\bsnm{Monette},~\bfnm{Georges}\binits{G.}} \AND
  \bauthor{\bsnm{Novomestky},~\bfnm{Frederick}\binits{F.}}
(\byear{2022}).
\btitle{Structural Equation Models}
\bnote{R package version 3.1-15}.
\end{bmanual}
\endbibitem

\bibitem[\protect\citeauthoryear{Friguet, Kloareg and
  Causeur}{2009}]{FriguetKloaregCauseur2009}
\begin{barticle}[author]
\bauthor{\bsnm{Friguet},~\bfnm{Chlo\'{e}}\binits{C.}},
  \bauthor{\bsnm{Kloareg},~\bfnm{Maela}\binits{M.}} \AND
  \bauthor{\bsnm{Causeur},~\bfnm{David}\binits{D.}}
(\byear{2009}).
\btitle{A factor model approach to multiple testing under dependence}.
\bjournal{J. Amer. Statist. Assoc.}
\bvolume{104}
\bpages{1406--1415}.
\bdoi{10.1198/jasa.2009.tm08332}
\bmrnumber{2750571}
\end{barticle}
\endbibitem

\bibitem[\protect\citeauthoryear{Gana and Broc}{2019}]{GanaBroc2019}
\begin{bbook}[author]
\bauthor{\bsnm{Gana},~\bfnm{Kamel}\binits{K.}} \AND
  \bauthor{\bsnm{Broc},~\bfnm{Guillaume}\binits{G.}}
(\byear{2019}).
\btitle{Structural equation modeling with lavaan}.
\bseries{Mathematics and Statistics Series}.
\bpublisher{ISTE, London; John Wiley \& Sons, Inc., Hoboken, NJ}.
\bmrnumber{3931337}
\end{bbook}
\endbibitem

\bibitem[\protect\citeauthoryear{Girvan and Newman}{2002}]{GirvanNewman2002}
\begin{barticle}[author]
\bauthor{\bsnm{Girvan},~\bfnm{M.}\binits{M.}} \AND
  \bauthor{\bsnm{Newman},~\bfnm{M.~E.~J.}\binits{M.~E.~J.}}
(\byear{2002}).
\btitle{Community structure in social and biological networks}.
\bjournal{Proc. Natl. Acad. Sci. USA}
\bvolume{99}
\bpages{7821--7826}.
\bdoi{10.1073/pnas.122653799}
\bmrnumber{1908073}
\end{barticle}
\endbibitem

\bibitem[\protect\citeauthoryear{Huttlin
  et~al.}{2017}]{HuttlinBrucknerPaulo2017}
\begin{barticle}[author]
\bauthor{\bsnm{Huttlin},~\bfnm{Edward~L}\binits{E.~L.}},
  \bauthor{\bsnm{Bruckner},~\bfnm{Raphael~J}\binits{R.~J.}},
  \bauthor{\bsnm{Paulo},~\bfnm{Joao~A}\binits{J.~A.}},
  \bauthor{\bsnm{Cannon},~\bfnm{Joe~R}\binits{J.~R.}},
  \bauthor{\bsnm{Ting},~\bfnm{Lily}\binits{L.}},
  \bauthor{\bsnm{Baltier},~\bfnm{Kurt}\binits{K.}},
  \bauthor{\bsnm{Colby},~\bfnm{Greg}\binits{G.}},
  \bauthor{\bsnm{Gebreab},~\bfnm{Fana}\binits{F.}},
  \bauthor{\bsnm{Gygi},~\bfnm{Melanie~P}\binits{M.~P.}},
  \bauthor{\bsnm{Parzen},~\bfnm{Hannah}\binits{H.}} \betal{et~al.}
(\byear{2017}).
\btitle{Architecture of the human interactome defines protein communities and
  disease networks}.
\bjournal{Nature}
\bvolume{545}
\bpages{505--509}.
\end{barticle}
\endbibitem

\bibitem[\protect\citeauthoryear{ISGlobal}{2021}]{ISG2021}
\begin{bmisc}[author]
\bauthor{\bsnm{ISGlobal}}
(\byear{2021}).
\btitle{Barcelona Institute for Global Health}.
\bhowpublished{\url{https://www.isglobal.org/en}}.
\bnote{Accessed: 2022-07-30}.
\end{bmisc}
\endbibitem

\bibitem[\protect\citeauthoryear{Jackson, Gillaspy~Jr and
  Purc-Stephenson}{2009}]{JacksonGillaspyJrPurcStephenson2009}
\begin{barticle}[author]
\bauthor{\bsnm{Jackson},~\bfnm{Dennis~L}\binits{D.~L.}},
  \bauthor{\bsnm{Gillaspy~Jr},~\bfnm{J~Arthur}\binits{J.~A.}} \AND
  \bauthor{\bsnm{Purc-Stephenson},~\bfnm{Rebecca}\binits{R.}}
(\byear{2009}).
\btitle{Reporting practices in confirmatory factor analysis: an overview and
  some recommendations}.
\bjournal{Psychological methods}
\bvolume{14}
\bpages{6}.
\end{barticle}
\endbibitem

\bibitem[\protect\citeauthoryear{J{\"o}reskog}{1969}]{Joreskog1969}
\begin{barticle}[author]
\bauthor{\bsnm{J{\"o}reskog},~\bfnm{Karl~G}\binits{K.~G.}}
(\byear{1969}).
\btitle{A general approach to confirmatory maximum likelihood factor analysis}.
\bjournal{Psychometrika}
\bvolume{34}
\bpages{183--202}.
\end{barticle}
\endbibitem

\bibitem[\protect\citeauthoryear{Lawley}{1958}]{Lawley1958}
\begin{barticle}[author]
\bauthor{\bsnm{Lawley},~\bfnm{DN}\binits{D.}}
(\byear{1958}).
\btitle{Estimation in factor analysis under various initial assumptions}.
\bjournal{British journal of statistical Psychology}
\bvolume{11}
\bpages{1--12}.
\end{barticle}
\endbibitem

\bibitem[\protect\citeauthoryear{Lei and Rinaldo}{2015}]{LeiRinaldo2015}
\begin{barticle}[author]
\bauthor{\bsnm{Lei},~\bfnm{Jing}\binits{J.}} \AND
  \bauthor{\bsnm{Rinaldo},~\bfnm{Alessandro}\binits{A.}}
(\byear{2015}).
\btitle{Consistency of spectral clustering in stochastic block models}.
\bjournal{Ann. Statist.}
\bvolume{43}
\bpages{215--237}.
\bdoi{10.1214/14-AOS1274}
\bmrnumber{3285605}
\end{barticle}
\endbibitem

\bibitem[\protect\citeauthoryear{Levine
  et~al.}{2015}]{LevineSimondsBendall2015}
\begin{barticle}[author]
\bauthor{\bsnm{Levine},~\bfnm{Jacob~H}\binits{J.~H.}},
  \bauthor{\bsnm{Simonds},~\bfnm{Erin~F}\binits{E.~F.}},
  \bauthor{\bsnm{Bendall},~\bfnm{Sean~C}\binits{S.~C.}},
  \bauthor{\bsnm{Davis},~\bfnm{Kara~L}\binits{K.~L.}},
  \bauthor{\bsnm{El-ad},~\bfnm{D~Amir}\binits{D.~A.}},
  \bauthor{\bsnm{Tadmor},~\bfnm{Michelle~D}\binits{M.~D.}},
  \bauthor{\bsnm{Litvin},~\bfnm{Oren}\binits{O.}},
  \bauthor{\bsnm{Fienberg},~\bfnm{Harris~G}\binits{H.~G.}},
  \bauthor{\bsnm{Jager},~\bfnm{Astraea}\binits{A.}},
  \bauthor{\bsnm{Zunder},~\bfnm{Eli~R}\binits{E.~R.}} \betal{et~al.}
(\byear{2015}).
\btitle{Data-driven phenotypic dissection of AML reveals progenitor-like cells
  that correlate with prognosis}.
\bjournal{Cell}
\bvolume{162}
\bpages{184--197}.
\end{barticle}
\endbibitem

\bibitem[\protect\citeauthoryear{Li et~al.}{2022}]{LiLeiBhattacharyya2022}
\begin{barticle}[author]
\bauthor{\bsnm{Li},~\bfnm{Tianxi}\binits{T.}},
  \bauthor{\bsnm{Lei},~\bfnm{Lihua}\binits{L.}},
  \bauthor{\bsnm{Bhattacharyya},~\bfnm{Sharmodeep}\binits{S.}},
  \bauthor{\bparticle{Van~den} \bsnm{Berge},~\bfnm{Koen}\binits{K.}},
  \bauthor{\bsnm{Sarkar},~\bfnm{Purnamrita}\binits{P.}},
  \bauthor{\bsnm{Bickel},~\bfnm{Peter~J.}\binits{P.~J.}} \AND
  \bauthor{\bsnm{Levina},~\bfnm{Elizaveta}\binits{E.}}
(\byear{2022}).
\btitle{Hierarchical community detection by recursive partitioning}.
\bjournal{J. Amer. Statist. Assoc.}
\bvolume{117}
\bpages{951--968}.
\bdoi{10.1080/01621459.2020.1833888}
\bmrnumber{4436325}
\end{barticle}
\endbibitem

\bibitem[\protect\citeauthoryear{Neale et~al.}{2016}]{NealeHunterPritikin2016}
\begin{barticle}[author]
\bauthor{\bsnm{Neale},~\bfnm{Michael~C.}\binits{M.~C.}},
  \bauthor{\bsnm{Hunter},~\bfnm{Michael~D.}\binits{M.~D.}},
  \bauthor{\bsnm{Pritikin},~\bfnm{Joshua~N.}\binits{J.~N.}},
  \bauthor{\bsnm{Zahery},~\bfnm{Mahsa}\binits{M.}},
  \bauthor{\bsnm{Brick},~\bfnm{Timothy~R.}\binits{T.~R.}},
  \bauthor{\bsnm{Kirkpatrick},~\bfnm{Robert~M.}\binits{R.~M.}},
  \bauthor{\bsnm{Estabrook},~\bfnm{Ryne}\binits{R.}},
  \bauthor{\bsnm{Bates},~\bfnm{Timothy~C.}\binits{T.~C.}},
  \bauthor{\bsnm{Maes},~\bfnm{Hermine~H.}\binits{H.~H.}} \AND
  \bauthor{\bsnm{Boker},~\bfnm{Steven~M.}\binits{S.~M.}}
(\byear{2016}).
\btitle{Open{M}x 2.0: extended structural equation and statistical modeling}.
\bjournal{Psychometrika}
\bvolume{81}
\bpages{535--549}.
\bdoi{10.1007/s11336-014-9435-8}
\bmrnumber{3505378}
\end{barticle}
\endbibitem

\bibitem[\protect\citeauthoryear{Newman and Girvan}{2004}]{NewmanGirvan2004}
\begin{barticle}[author]
\bauthor{\bsnm{Newman},~\bfnm{M.~E.~J.}\binits{M.~E.~J.}} \AND
  \bauthor{\bsnm{Girvan},~\bfnm{M.}\binits{M.}}
(\byear{2004}).
\btitle{Finding and evaluating community structure in networks}.
\bjournal{Phys. Rev. E}
\bvolume{69}
\bpages{026113}.
\bdoi{10.1103/PhysRevE.69.026113}
\end{barticle}
\endbibitem

\bibitem[\protect\citeauthoryear{Oberski}{2014}]{Oberski2014}
\begin{barticle}[author]
\bauthor{\bsnm{Oberski},~\bfnm{Daniel}\binits{D.}}
(\byear{2014}).
\btitle{lavaan.survey: An R Package for Complex Survey Analysis of Structural
  Equation Models}.
\bjournal{Journal of Statistical Software}
\bvolume{57}
\bpages{1-–27}.
\bdoi{10.18637/jss.v057.i01}
\end{barticle}
\endbibitem

\bibitem[\protect\citeauthoryear{Perrot-Dock\`es, L\'{e}vy-Leduc and
  Rajjou}{2022}]{PerrotLevyRajjou2022}
\begin{barticle}[author]
\bauthor{\bsnm{Perrot-Dock\`es},~\bfnm{M.}\binits{M.}},
  \bauthor{\bsnm{L\'{e}vy-Leduc},~\bfnm{C.}\binits{C.}} \AND
  \bauthor{\bsnm{Rajjou},~\bfnm{L.}\binits{L.}}
(\byear{2022}).
\btitle{Estimation of large block structured covariance matrices: application
  to `multi-omic' approaches to study seed quality}.
\bjournal{J. R. Stat. Soc. Ser. C. Appl. Stat.}
\bvolume{71}
\bpages{119--147}.
\bdoi{10.1111/rssc.12524}
\bmrnumber{4376848}
\end{barticle}
\endbibitem

\bibitem[\protect\citeauthoryear{Ritchie
  et~al.}{2023}]{RitchieSurendranKarthikeyan2023}
\begin{barticle}[author]
\bauthor{\bsnm{Ritchie},~\bfnm{Scott~C}\binits{S.~C.}},
  \bauthor{\bsnm{Surendran},~\bfnm{Praveen}\binits{P.}},
  \bauthor{\bsnm{Karthikeyan},~\bfnm{Savita}\binits{S.}},
  \bauthor{\bsnm{Lambert},~\bfnm{Samuel~A}\binits{S.~A.}},
  \bauthor{\bsnm{Bolton},~\bfnm{Thomas}\binits{T.}},
  \bauthor{\bsnm{Pennells},~\bfnm{Lisa}\binits{L.}},
  \bauthor{\bsnm{Danesh},~\bfnm{John}\binits{J.}},
  \bauthor{\bsnm{Di~Angelantonio},~\bfnm{Emanuele}\binits{E.}},
  \bauthor{\bsnm{Butterworth},~\bfnm{Adam~S}\binits{A.~S.}} \AND
  \bauthor{\bsnm{Inouye},~\bfnm{Michael}\binits{M.}}
(\byear{2023}).
\btitle{Quality control and removal of technical variation of NMR metabolic
  biomarker data in\~{} 120,000 UK Biobank participants}.
\bjournal{Scientific Data}
\bvolume{10}
\bpages{64}.
\end{barticle}
\endbibitem

\bibitem[\protect\citeauthoryear{Rosseel}{2012}]{Rosseel2012}
\begin{barticle}[author]
\bauthor{\bsnm{Rosseel},~\bfnm{Yves}\binits{Y.}}
(\byear{2012}).
\btitle{lavaan: An R Package for Structural Equation Modeling}.
\bjournal{Journal of Statistical Software}
\bvolume{48}
\bpages{1–-36}.
\bdoi{10.18637/jss.v048.i02}
\end{barticle}
\endbibitem

\bibitem[\protect\citeauthoryear{Rosseel
  et~al.}{2023}]{RRosseelJorgensenRockwood2023}
\begin{bmanual}[author]
\bauthor{\bsnm{Rosseel},~\bfnm{Yves}\binits{Y.}},
  \bauthor{\bsnm{Jorgensen},~\bfnm{Terrence~D.}\binits{T.~D.}},
  \bauthor{\bsnm{Rockwood},~\bfnm{Nicholas}\binits{N.}},
  \bauthor{\bsnm{Oberski},~\bfnm{Daniel}\binits{D.}},
  \bauthor{\bsnm{Byrnes},~\bfnm{Jarrett}\binits{J.}},
  \bauthor{\bsnm{Vanbrabant},~\bfnm{Leonard}\binits{L.}},
  \bauthor{\bsnm{Savalei},~\bfnm{Victoria}\binits{V.}},
  \bauthor{\bsnm{Merkle},~\bfnm{Ed}\binits{E.}},
  \bauthor{\bsnm{Hallquist},~\bfnm{Michael}\binits{M.}},
  \bauthor{\bsnm{Rhemtulla},~\bfnm{Mijke}\binits{M.}},
  \bauthor{\bsnm{Katsikatsou},~\bfnm{Myrsini}\binits{M.}},
  \bauthor{\bsnm{Barendse},~\bfnm{Mariska}\binits{M.}},
  \bauthor{\bsnm{Scharf},~\bfnm{Florian}\binits{F.}} \AND
  \bauthor{\bsnm{Du},~\bfnm{Han}\binits{H.}}
(\byear{2023}).
\btitle{Latent Variable Analysis}
\bnote{R package version 0.6-15}.
\end{bmanual}
\endbibitem

\bibitem[\protect\citeauthoryear{Schaub, Li and Peel}{2023}]{SchaubLiPeel2023}
\begin{barticle}[author]
\bauthor{\bsnm{Schaub},~\bfnm{Michael~T.}\binits{M.~T.}},
  \bauthor{\bsnm{Li},~\bfnm{Jiaze}\binits{J.}} \AND
  \bauthor{\bsnm{Peel},~\bfnm{Leto}\binits{L.}}
(\byear{2023}).
\btitle{Hierarchical community structure in networks}.
\bjournal{Phys. Rev. E}
\bvolume{107}
\bpages{Paper No. 054305, 22}.
\bdoi{10.1103/physreve.107.054305}
\bmrnumber{4606010}
\end{barticle}
\endbibitem

\bibitem[\protect\citeauthoryear{Schreiber
  et~al.}{2006}]{SchreiberNoraStage2006}
\begin{barticle}[author]
\bauthor{\bsnm{Schreiber},~\bfnm{James~B.}\binits{J.~B.}},
  \bauthor{\bsnm{Nora},~\bfnm{Amaury}\binits{A.}},
  \bauthor{\bsnm{Stage},~\bfnm{Frances~K.}\binits{F.~K.}},
  \bauthor{\bsnm{Barlow},~\bfnm{Elizabeth~A.}\binits{E.~A.}} \AND
  \bauthor{\bsnm{King},~\bfnm{Jamie}\binits{J.}}
(\byear{2006}).
\btitle{Reporting Structural Equation Modeling and Confirmatory Factor Analysis
  Results: A Review}.
\bjournal{The Journal of Educational Research}
\bvolume{99}
\bpages{323--338}.
\bdoi{10.3200/JOER.99.6.323-338}
\end{barticle}
\endbibitem

\bibitem[\protect\citeauthoryear{Simpson, Bowman and
  Laurienti}{2013}]{SimpsonBowmanLaurienti2013}
\begin{barticle}[author]
\bauthor{\bsnm{Simpson},~\bfnm{Sean~L.}\binits{S.~L.}},
  \bauthor{\bsnm{Bowman},~\bfnm{F.~DuBois}\binits{F.~D.}} \AND
  \bauthor{\bsnm{Laurienti},~\bfnm{Paul~J.}\binits{P.~J.}}
(\byear{2013}).
\btitle{Analyzing complex functional brain networks: fusing statistics and
  network science to understand the brain}.
\bjournal{Stat. Surv.}
\bvolume{7}
\bpages{1--36}.
\bdoi{10.1214/13-SS103}
\bmrnumber{3161730}
\end{barticle}
\endbibitem

\bibitem[\protect\citeauthoryear{Tomczak, Czerwi{\'n}ska and
  Wiznerowicz}{2015}]{KatarzynaPatrycjaMaciej2015}
\begin{barticle}[author]
\bauthor{\bsnm{Tomczak},~\bfnm{Katarzyna}\binits{K.}},
  \bauthor{\bsnm{Czerwi{\'n}ska},~\bfnm{Patrycja}\binits{P.}} \AND
  \bauthor{\bsnm{Wiznerowicz},~\bfnm{Maciej}\binits{M.}}
(\byear{2015}).
\btitle{Review The Cancer Genome Atlas (TCGA): an immeasurable source of
  knowledge}.
\bjournal{Contemporary Oncology/Wsp{\'o}{\l}czesna Onkologia}
\bvolume{2015}
\bpages{68--77}.
\end{barticle}
\endbibitem

\bibitem[\protect\citeauthoryear{Wallace}{2012}]{Wallace2012}
\begin{barticle}[author]
\bauthor{\bsnm{Wallace},~\bfnm{Douglas~C}\binits{D.~C.}}
(\byear{2012}).
\btitle{Mitochondria and cancer}.
\bjournal{Nature Reviews Cancer}
\bvolume{12}
\bpages{685--698}.
\end{barticle}
\endbibitem

\bibitem[\protect\citeauthoryear{Wang, Liang and Ji}{2020}]{WangLiangJi2020}
\begin{barticle}[author]
\bauthor{\bsnm{Wang},~\bfnm{Zhe}\binits{Z.}},
  \bauthor{\bsnm{Liang},~\bfnm{Yingbin}\binits{Y.}} \AND
  \bauthor{\bsnm{Ji},~\bfnm{Pengsheng}\binits{P.}}
(\byear{2020}).
\btitle{Spectral algorithms for community detection in directed networks}.
\bjournal{J. Mach. Learn. Res.}
\bvolume{21}
\bpages{Paper No. 153, 45}.
\bmrnumber{4209439}
\end{barticle}
\endbibitem

\bibitem[\protect\citeauthoryear{Wu et~al.}{2021}]{WuMaLiu2021}
\begin{barticle}[author]
\bauthor{\bsnm{Wu},~\bfnm{Qiong}\binits{Q.}},
  \bauthor{\bsnm{Ma},~\bfnm{Tianzhou}\binits{T.}},
  \bauthor{\bsnm{Liu},~\bfnm{Qingzhi}\binits{Q.}},
  \bauthor{\bsnm{Milton},~\bfnm{Donald~K}\binits{D.~K.}},
  \bauthor{\bsnm{Zhang},~\bfnm{Yuan}\binits{Y.}} \AND
  \bauthor{\bsnm{Chen},~\bfnm{Shuo}\binits{S.}}
(\byear{2021}).
\btitle{ICN: extracting interconnected communities in gene co-expression
  networks}.
\bjournal{Bioinformatics}
\bvolume{37}
\bpages{1997--2003}.
\bdoi{10.1093/bioinformatics/btab047}
\end{barticle}
\endbibitem

\bibitem[\protect\citeauthoryear{Yang, Chen and Chen}{2024}]{YangChenChen2024}
\begin{barticle}[author]
\bauthor{\bsnm{Yang},~\bfnm{Yifan}\binits{Y.}},
  \bauthor{\bsnm{Chen},~\bfnm{Chixiang}\binits{C.}} \AND
  \bauthor{\bsnm{Chen},~\bfnm{Shuo}\binits{S.}}
(\byear{2024}).
\btitle{Covariance matrix estimation for high-throughput biomedical data with
  interconnected communities}.
\bjournal{The American Statistician}.
\end{barticle}
\endbibitem

\bibitem[\protect\citeauthoryear{Zitnik, Sosi{\u{c}} and
  Leskovec}{2018}]{ZitnikSosicLeskovec2018}
\begin{barticle}[author]
\bauthor{\bsnm{Zitnik},~\bfnm{Marinka}\binits{M.}},
  \bauthor{\bsnm{Sosi{\u{c}}},~\bfnm{Rok}\binits{R.}} \AND
  \bauthor{\bsnm{Leskovec},~\bfnm{Jure}\binits{J.}}
(\byear{2018}).
\btitle{Prioritizing network communities}.
\bjournal{Nature Communications}
\bvolume{9}
\bpages{2544}.
\end{barticle}
\endbibitem

\end{thebibliography}


\begin{thebibliography}{6}

\bibitem[\protect\citeauthoryear{Lei and Rinaldo}{2015}]{LeiRinaldo2015}
\begin{barticle}[author]
\bauthor{\bsnm{Lei},~\bfnm{Jing}\binits{J.}} \AND
  \bauthor{\bsnm{Rinaldo},~\bfnm{Alessandro}\binits{A.}}
(\byear{2015}).
\btitle{Consistency of spectral clustering in stochastic block models}.
\bjournal{Ann. Statist.}
\bvolume{43}
\bpages{215--237}.
\bdoi{10.1214/14-AOS1274}
\bmrnumber{3285605}
\end{barticle}
\endbibitem

\bibitem[\protect\citeauthoryear{Li et~al.}{2022}]{LiLeiBhattacharyya2022}
\begin{barticle}[author]
\bauthor{\bsnm{Li},~\bfnm{Tianxi}\binits{T.}},
  \bauthor{\bsnm{Lei},~\bfnm{Lihua}\binits{L.}},
  \bauthor{\bsnm{Bhattacharyya},~\bfnm{Sharmodeep}\binits{S.}},
  \bauthor{\bparticle{Van~den} \bsnm{Berge},~\bfnm{Koen}\binits{K.}},
  \bauthor{\bsnm{Sarkar},~\bfnm{Purnamrita}\binits{P.}},
  \bauthor{\bsnm{Bickel},~\bfnm{Peter~J.}\binits{P.~J.}} \AND
  \bauthor{\bsnm{Levina},~\bfnm{Elizaveta}\binits{E.}}
(\byear{2022}).
\btitle{Hierarchical community detection by recursive partitioning}.
\bjournal{J. Amer. Statist. Assoc.}
\bvolume{117}
\bpages{951--968}.
\bdoi{10.1080/01621459.2020.1833888}
\bmrnumber{4436325}
\end{barticle}
\endbibitem

\bibitem[\protect\citeauthoryear{Lu and Schmidt}{2012}]{LuSchmidt2012}
\begin{barticle}[author]
\bauthor{\bsnm{Lu},~\bfnm{Cuicui}\binits{C.}} \AND
  \bauthor{\bsnm{Schmidt},~\bfnm{Peter}\binits{P.}}
(\byear{2012}).
\btitle{Conditions for the numerical equality of the {OLS}, {GLS} and
  {A}memiya-{C}ragg estimators}.
\bjournal{Econom. Lett.}
\bvolume{116}
\bpages{538--540}.
\bdoi{10.1016/j.econlet.2012.01.015}
\bmrnumber{2965868}
\end{barticle}
\endbibitem

\bibitem[\protect\citeauthoryear{Wu et~al.}{2021}]{WuMaLiu2021}
\begin{barticle}[author]
\bauthor{\bsnm{Wu},~\bfnm{Qiong}\binits{Q.}},
  \bauthor{\bsnm{Ma},~\bfnm{Tianzhou}\binits{T.}},
  \bauthor{\bsnm{Liu},~\bfnm{Qingzhi}\binits{Q.}},
  \bauthor{\bsnm{Milton},~\bfnm{Donald~K}\binits{D.~K.}},
  \bauthor{\bsnm{Zhang},~\bfnm{Yuan}\binits{Y.}} \AND
  \bauthor{\bsnm{Chen},~\bfnm{Shuo}\binits{S.}}
(\byear{2021}).
\btitle{ICN: extracting interconnected communities in gene co-expression
  networks}.
\bjournal{Bioinformatics}
\bvolume{37}
\bpages{1997--2003}.
\bdoi{10.1093/bioinformatics/btab047}
\end{barticle}
\endbibitem

\bibitem[\protect\citeauthoryear{Yang, Chen and Chen}{2024}]{YangChenChen2024}
\begin{barticle}[author]
\bauthor{\bsnm{Yang},~\bfnm{Yifan}\binits{Y.}},
  \bauthor{\bsnm{Chen},~\bfnm{Chixiang}\binits{C.}} \AND
  \bauthor{\bsnm{Chen},~\bfnm{Shuo}\binits{S.}}
(\byear{2024}).
\btitle{Covariance matrix estimation for high-throughput biomedical data with
  interconnected communities}.
\bjournal{The American Statistician}.
\end{barticle}
\endbibitem

\bibitem[\protect\citeauthoryear{Yang et~al.}{2024}]{YangChenLee2024}
\begin{bmisc}[author]
\bauthor{\bsnm{Yang},~\bfnm{Yifan}\binits{Y.}},
  \bauthor{\bsnm{Chen},~\bfnm{Shuo}\binits{S.}},
  \bauthor{\bsnm{Lee},~\bfnm{Hwiyoung}\binits{H.}} \AND
  \bauthor{\bsnm{Wang},~\bfnm{Ming}\binits{M.}}
(\byear{2024}).
\btitle{A New Representation of Uniform-Block Matrix and Applications}.
\end{bmisc}
\endbibitem

\end{thebibliography}


\end{document}


\begin{frontmatter}
		\title{Supplementary Material For Semi-Confirmatory Factor Analysis for High-Dimensional Data with Interconnected Community Structures}
		\runtitle{SCFA}
		
		\begin{aug}
			\author[A]{\fnms{Yifan}~\snm{Yang}\ead[label=e1]{yiorfun@case.edu} \orcid{0000-0001-5727-6540}},
			\author[B]{\fnms{Tianzhou}~\snm{Ma}\ead[label=e2]{tma0929@umd.edu}}, 
			\author[C]{\fnms{Chuan}~\snm{Bi}\ead[label=e3]{Chuan.Bi@som.umaryland.edu}}
			\and
			\author[C]{\fnms{Shuo}~\snm{Chen}\ead[label=e4]{shuochen@som.umaryland.edu}}
			\address[A]{Department of Population and Quantitative Health Sciences, Case Western Reserve University \printead[presep={,\ }]{e1}}
			
			\address[B]{Department of Epidemiology and Biostatistics, University of Maryland, College Park \printead[presep={,\ }]{e2}}
			
			\address[C]{School of Medicine, University of Maryland \printead[presep={,\ }]{e3,e4}}
		\end{aug}
		
		\begin{abstract}
			The present supplementary material contains four sections. 
			Section~\ref{Sec:ub_definitions} contains the definitions related to uniform-block matrix and uniform-block structure.
			Section~\ref{Sec:ub_properties} includes the properties of the uniform-block matrices.
			Section~\ref{Sec:numerical} has the additional numerical results for both Simulation Studies and Applications to Gene Expression Data sections. 
			Section~\ref{Sec:simulation} presents the additional simulation studies.
			Section~\ref{Sec:proof} provides the technical proofs.
			In addition to this document, the online material folder contains some additional simulation results presented in Excel files and the R code used for simulation studies in Section 3 and applications to gene expression data in Section 4. 
			The code is accessible at \url{https://github.com/yiorfun/SCFA}.
			An R package is also available at \url{https://github.com/xavienzo/SCFA}.
		\end{abstract}

	\end{frontmatter}
	

	
	We first introduce some fundamental concepts related to the interconnected community structure and then establish the inference results for SCFA models.
	
	\section{DEFINITIONS}
	
	\label{Sec:ub_definitions}
	
	\begin{dfns}[partition-size vector and partitioned matrix by a partition-size vector]
		\label{Dfn:par}
		Suppose $\bm{p} = \left(p_1, \ldots, p_K\right) ^ \top \in \mathbb{Z}_{+} ^ {K \times 1}$ is a column vector and 
		matrix $\textbf{N} \in \mathbb{R} ^ {p \times p}$ has a $K$ by $K$ blocks form $\left(\textbf{N}_{kk'}\right)$. 
		Then, 
		
		(1) $\bm{p}$ is said to be a partition-size vector, if $p = p_1 + \cdots + p_K$ and $p_k > 1$ for all $k$; 
		
		(2) $\left(\textbf{N}_{kk'}\right)$ is said to be the partitioned matrix of $\textbf{N}$ by $\bm{p}$, if the block $\textbf{N}_{kk'}$ has dimensions $p_k$ by $p_{k'}$ for $k, k' = 1, \ldots, K$.
	\end{dfns}
	
	\begin{dfns}[uniform-block matrix and uniform-block structure]
		\label{Dfn:ub}
		Suppose $\bm{p} = \left(p_1, \ldots, p_K\right) ^ \top$ is a partition-size vector satisfying $p = p_1 + \cdots + p_K$ and $p_k > 1$ for all $k$, 
		and $\left(\textbf{N}_{kk'}\right)$ is the partitioned matrix of symmetric $\textbf{N} \in \mathbb{R} ^ {p \times p}$ by $\bm{p}$.
		Then, $\left(\textbf{N}_{kk'}\right)$ is said to be a uniform-block matrix, if there exist $a_{kk}, b_{kk'} \in \mathbb{R}$ satisfying the diagonal block $\textbf{N}_{kk} = a_{kk} \textbf{I}_{p_k} + b_{kk} \textbf{J}_{p_k}$ for $k = k'$, the off-diagonal block $\textbf{N}_{kk'} = b_{kk'} \textbf{1}_{p_k \times p_{k'}}$ with $b_{k'k} = b_{kk'}$ for $k \neq k'$; 		
		the structure of a uniform-block matrix is said to be a uniform-block structure.		
		We denote the above uniform-block matrix by $\left(\textbf{N}_{kk'}\right) = \textbf{N}\left(\textbf{A}, \textbf{B}, \bm{p} \right)$,
		where $\textbf{A} = \operatorname{diag}\left(a_{11}, \ldots, a_{KK}\right)$ is a $K$ by $K$ diagonal matrix and $\textbf{B} = \left(b_{kk'}\right)$ is a $K$ by $K$ symmetric matrix with $b_{kk'} = b_{k'k}$ for $k \neq k'$.
	\end{dfns}
	
	Following Definition~\ref{Dfn:ub}, the covariance matrix with an interconnected community structure defined as (3) in the manuscript, i.e., 
	\begin{equation*}
		\bm{\Sigma}_{p \times p} 
		= \left(\bm{\Sigma}_{kk'}\right) 
		= \begin{pmatrix}
			a_{11} \textbf{I}_{p_1} + b_{11} \textbf{J}_{p_1} & b_{12} \textbf{1}_{p_1 \times p_2} & \cdots & b_{1K} \textbf{1}_{p_1 \times p_K} \\
			b_{21} \textbf{1}_{p_2 \times p_1} & a_{22} \textbf{I}_{p_2} + b_{22} \textbf{J}_{p_2} & \cdots & b_{2K} \textbf{1}_{p_2 \times p_K} \\ 
			\vdots & \vdots & \ddots & \vdots \\
			b_{K1} \textbf{1}_{p_K \times p_1} & b_{K2} \textbf{1}_{p_K \times p_2} & \cdots & a_{KK} \textbf{I}_{p_K} + b_{KK} \textbf{J}_{p_K} 
		\end{pmatrix}, b_{kk'} = b_{k'k},
	\end{equation*}
	is a uniform-block matrix.
	More specifically, given a $K$-dimensional partition-size vector $\bm{p} = \left(p_1, \ldots, p_K \right) ^ \top$ satisfying that $p_k > 1$ and $p_1 + \cdots + p_K = p$, the partitioned matrix of $\bm{\Sigma}_{p \times p}$ by $\bm{p}$ is a uniform-block matrix $\bm{\Sigma} = \bm{\Sigma}\left(\textbf{A}, \textbf{B}, \bm{p}\right)$, 
	where $\textbf{A} = \operatorname{diag}\left(a_{11}, \ldots, a_{KK}\right)$ and $\textbf{B} = \left(b_{kk'}\right)$ with $b_{kk'} = b_{k'k}$ for $k \neq k'$.

	\section{PROPERTIES OF UNIFORM-BLOCK MATRICES}
	
	\label{Sec:ub_properties}
	
	\begin{cors}[block Hadamard product representation of a uniform-block matrix]
		\label{Cor:ub_rep}
		Suppose $\bm{p} = \left(p_1, \dots, p_K\right) ^ \top$ is a partition-size vector satisfying $p = p_1 + \cdots + p_K$ and $p_k > 1$ for all $k$, and $\textbf{N} \left(\textbf{A}, \textbf{B}, \bm{p} \right)$ is a uniform-block matrix with diagonal matrix $\textbf{A} = \operatorname{diag}\left(a_{11}, \ldots, a_{KK}\right)$ and symmetric matrix $\textbf{B} = \left(b_{k k'}\right)$ with $b_{kk'} = b_{k'k}$ for $k \neq k'$.
		Then, $\textbf{N} \left(\textbf{A}, \textbf{B}, \bm{p} \right)$ can be uniquely expressed by the sum of two block Hadamard products below: 
		\begin{equation*}
			\textbf{N}\left(\textbf{A}, \textbf{B}, \bm{p} \right) 
			= \textbf{A} \circ \textbf{I}(\bm{p}) + \textbf{B} \circ \textbf{J}(\bm{p}),
		\end{equation*}
		where $\textbf{I}(\bm{p})$ is a short notation for the ``identity'' uniform-block matrix $\textbf{I} \left(\textbf{I}_K, \textbf{0}_{K \times K}, \bm{p} \right) = \operatorname{Bdiag}\left(\textbf{I}_{p_1}, \ldots, \textbf{I}_{p_K}\right)$, $\textbf{J}(\bm{p})$ is a short notation for the ``all-one'' uniform-block matrix $\textbf{J} \left(\textbf{0}_{K \times K}, \textbf{J}_{K}, \bm{p} \right) = \left(\textbf{1}_{p_k \times p_{k'}}\right)$,
		and $\circ$ is the block Hadamard product: $\textbf{A} \circ \textbf{I}(\bm{p})$ equals $\operatorname{Bdiag}\left(a_{11} \textbf{I}_{p_1}, \ldots, a_{KK} \textbf{I}_{p_K}\right)$ and $\textbf{B} \circ \textbf{J}(\bm{p})$ equals $\left(b_{kk'} \textbf{1}_{p_k \times p_{k'}}\right)$, respectively.
	\end{cors}
	
	Due to the block Hadamard product representation of a uniform-block matrix, we may call diagonal matrix $\textbf{A}$, symmetric matrix $\textbf{B}$, and partition-size vector $\bm{p}$ as the ``coordinate matrices'' of a uniform-block matrix $\textbf{N}\left(\textbf{A}, \textbf{B}, \bm{p} \right)$.
	
	\begin{cors}[operations of uniform-block matrices]
		
		\label{Cor:ops_all}
		
		Given uniform-block matrices $\textbf{N} = \textbf{N}\left(\textbf{A}, \textbf{B}, \bm{p} \right)$, $\textbf{N}_1 = \textbf{N}_1\left(\textbf{A}_{1}, \textbf{B}_{1}, \bm{p} \right)$ and $\textbf{N}_2 = \textbf{N}_2 \left(\textbf{A}_{2}, \textbf{B}_{2}, \bm{p} \right)$ with diagonal matrices $\textbf{A} = \operatorname{diag}\left(a_{11}, \dots, a_{KK}\right)$, $\textbf{A}_{1}$, $\textbf{A}_{2}$, symmetric matrices $\textbf{B} = \left(b_{kk'}\right)$, $\textbf{B}_{1}$, $\textbf{B}_{2}$, and a common partition-size vector $\bm{p} = \left(p_1, \ldots, p_K\right) ^ \top$ satisfying $p_k > 1$ for every $k$ and $p = p_1 + \cdots + p_K$, respectively,
		let $\bm{\Delta} = \textbf{A} + \textbf{B} \textbf{P}$, 
		where $\textbf{P} = \operatorname{diag}(p_1, \ldots, p_K)$ is a $K$ by $K$ diagonal matrix. 
		
		(1) (Addition/Subtraction) suppose $\textbf{N} ^ * = \textbf{N}_1 \pm \textbf{N}_2$, then the partitioned matrix of $\textbf{N} ^ *$ by $\bm{p}$ is a uniform-block matrix, denoted by $\textbf{N} ^ * \left(\textbf{A}^*, \textbf{B}^*, \bm{p} \right)$, where $\textbf{A}^* = \textbf{A}_1 \pm \textbf{A}_2$ and $\textbf{B}^* = \textbf{B}_1 \pm \textbf{B}_2$;
		
		(2) (Square) suppose $\textbf{N}^* = \textbf{N} \times \textbf{N}$, then the partitioned matrix of $\textbf{N} ^ *$ by $\bm{p}$ is a uniform-block matrix, denoted by $\textbf{N} ^ *\left(\textbf{A}^*, \textbf{B}^*, \bm{p} \right)$, where $\textbf{A} ^ *  = \textbf{A} ^ 2$ and $\textbf{B}^* = \textbf{A} \textbf{B} + \textbf{B} \textbf{A} + \textbf{B} \textbf{P} \textbf{B}$;
		
		(3) (Eigenvalues) $\textbf{N}\left(\textbf{A}, \textbf{B}, \bm{p} \right)$ has $p$ eigenvalues in total, 
		those are $a_{kk}$ with multiplicity $(p_k - 1)$ for $k = 1, \dots, K$ and the rest $K$ eigenvalues are identical with those of $\bm{\Delta}$;
		
		(4) (Determinant) $\textbf{N}\left(\textbf{A}, \textbf{B}, \bm{p} \right)$ has the determinant $\left( \prod_{k = 1}^{K} a_{kk} ^ {p_k - 1} \right) \det\left( \bm{\Delta} \right)$; 
		
		(5) (Inverse) suppose $\textbf{N}$ is invertible and $\textbf{N}^* = \textbf{N} ^ {-1}$, then the partitioned matrix of $\textbf{N}^*$ by $\bm{p}$ is a uniform-block matrix, denoted by $\textbf{N} ^ *\left(\textbf{A}^*, \textbf{B}^*, \bm{p} \right)$, where $\textbf{A}^* = \textbf{A} ^ {- 1}$ and $\textbf{B} ^ * = - \bm{\Delta} ^ {- 1} \textbf{B} \textbf{A} ^ {- 1}$.
	\end{cors}
	
	\section{ADDITIONAL NUMERICAL RESULTS}
	
	\label{Sec:numerical}
	
	\subsection{Simulation studies}
	
	We conducted a comparative analysis between our proposed method and existing computational packages, namely ``sem'', ``OpenMx'', and ``lavaan'', focusing on the estimation of factor scores, the loading matrix $\textbf{L}$, the covariance matrix of common factors $\bm{\Sigma}_{\bm{f}}$, and the covariance matrix of errors $\bm{\Sigma}_{\bm{u}}$.
	Furthermore, we explored misspecification scenarios where the interconnected community structure is violated for the covariance matrix of observed variables $\bm{\Sigma}$.
	
	The simulation studies section presents a comparison of estimated factor scores and the finite-sample performance of estimators for $a_{kk}$ and $b_{kk'}$. 
	Additionally, estimation results for the misspecification analysis are reported in this section.
	
	The remanding results, including estimation results for the loading matrix $\textbf{L}$, all entries in $\bm{\Sigma}_{\bm{f}}$, and the diagonal entries in $\bm{\Sigma}_{\bm{u}}$ using the packages ``sem'', ``OpenMx'', and ``lavaan'',
	are provided in an Excel file available in the online material. 
	
	\subsection{Applications to gene expression data}
	
	We employed the SCFA model on both TCGA kidney cancer and GBM datasets and obtained estimation results for $a_{kk}$ and $b_{kk'}$, factor scores, and the membership of genes.
	Additionally, pathway analysis results of $6$ communities for kidney cancer dataset and $9$ communities for GBM dataset are included.
	These Excel files can be accessed in the online material.
	
	A complete version of the path diagram for the GBM dataset is provided below. 
	
	\begin{figure}[!h]
		\begin{center}
			\includegraphics[width = 0.7 \linewidth]{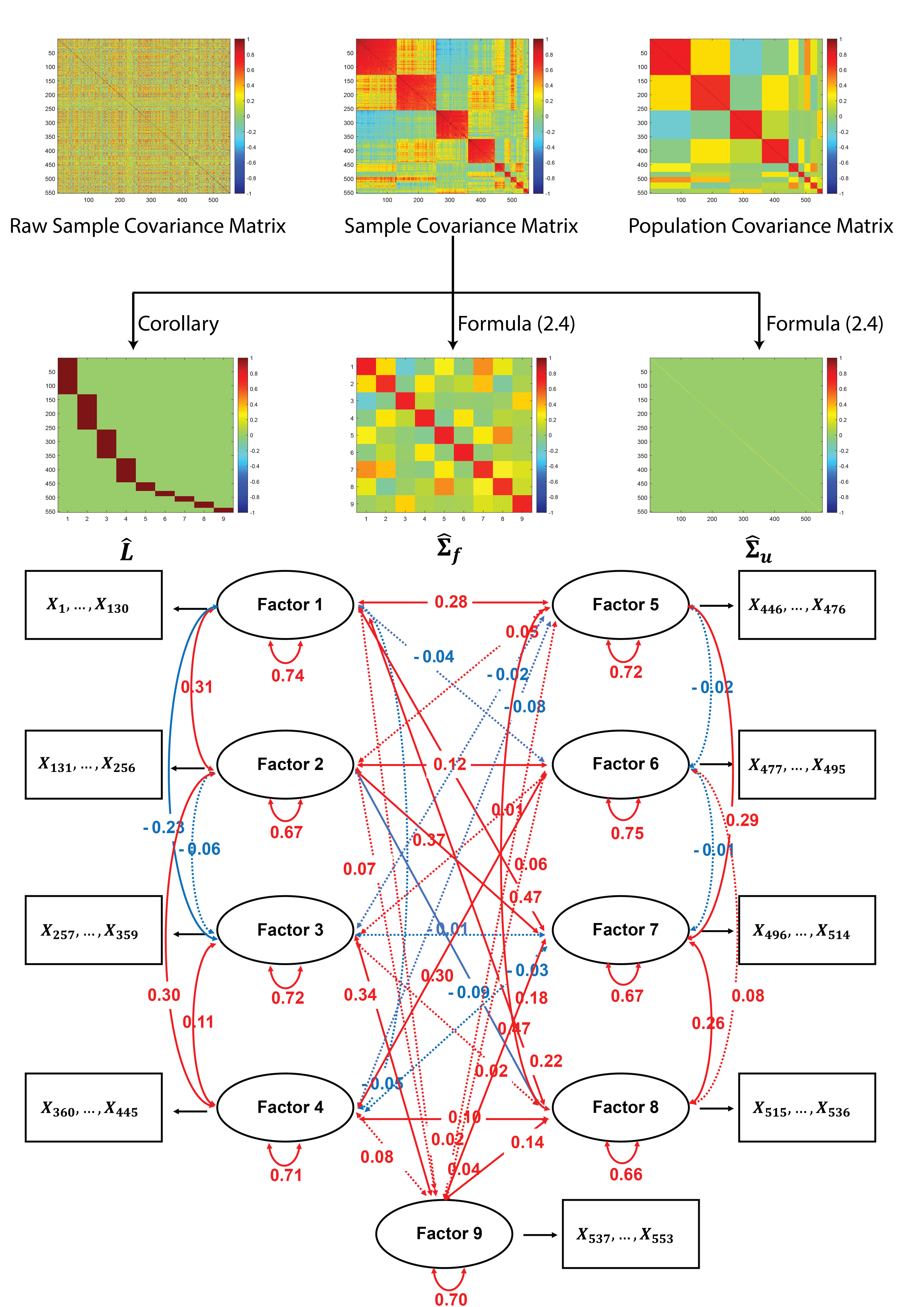} 
			\caption{
				The workflow (top) and complete version of the path diagram (bottom) for analyzing the TCGA GBM dataset.
				The subfigures in the first row of the workflow: 
				the input correlation matrix of $553$ genes (left), the reordered correlation matrix highlighting the
				detected interconnected community structure (middle), and the (estimated) population correlation matrix (right),
				respectively. 
				In the second row: the estimates $\widehat{\textbf{L}}$ (left), $\widehat{\bm{\Sigma}}_{\bm{f}}$ (middle), and $\widehat{\bm{\Sigma}}_{\bm{u}}$ (right), respectively. 
				The bottom path diagram illustrates gene subsets (in rectangles) and their respective
				common factors (in oval circles): 
				the lines (with double-ended arrows) connecting the $9$ common factors to themselves denote their
				estimated correlation coefficients (i.e., the diagonal entries of $\widehat{\bm{\Sigma}}_{\bm{f}}$);
				and the lines (with double-ended arrows) among the $9$ common factors denote their
				estimated correlation coefficients (i.e., the off-diagonal entries of $\widehat{\bm{\Sigma}}_{\bm{f}}$). Specifically, the red solid lines represent
				significant positive correlations, the blue solid lines represent significant negative correlations, and the dashed
				lines represent non-significant correlations.
			}
			\label{Fig:GBM}
		\end{center}
	\end{figure}
	
	\section{ADDITIONAL SIMULATION STUDIES}
	
	\label{Sec:simulation}
	
\begin{figure}[!h]
		\begin{center}
			\includegraphics[width = 1.0\linewidth]{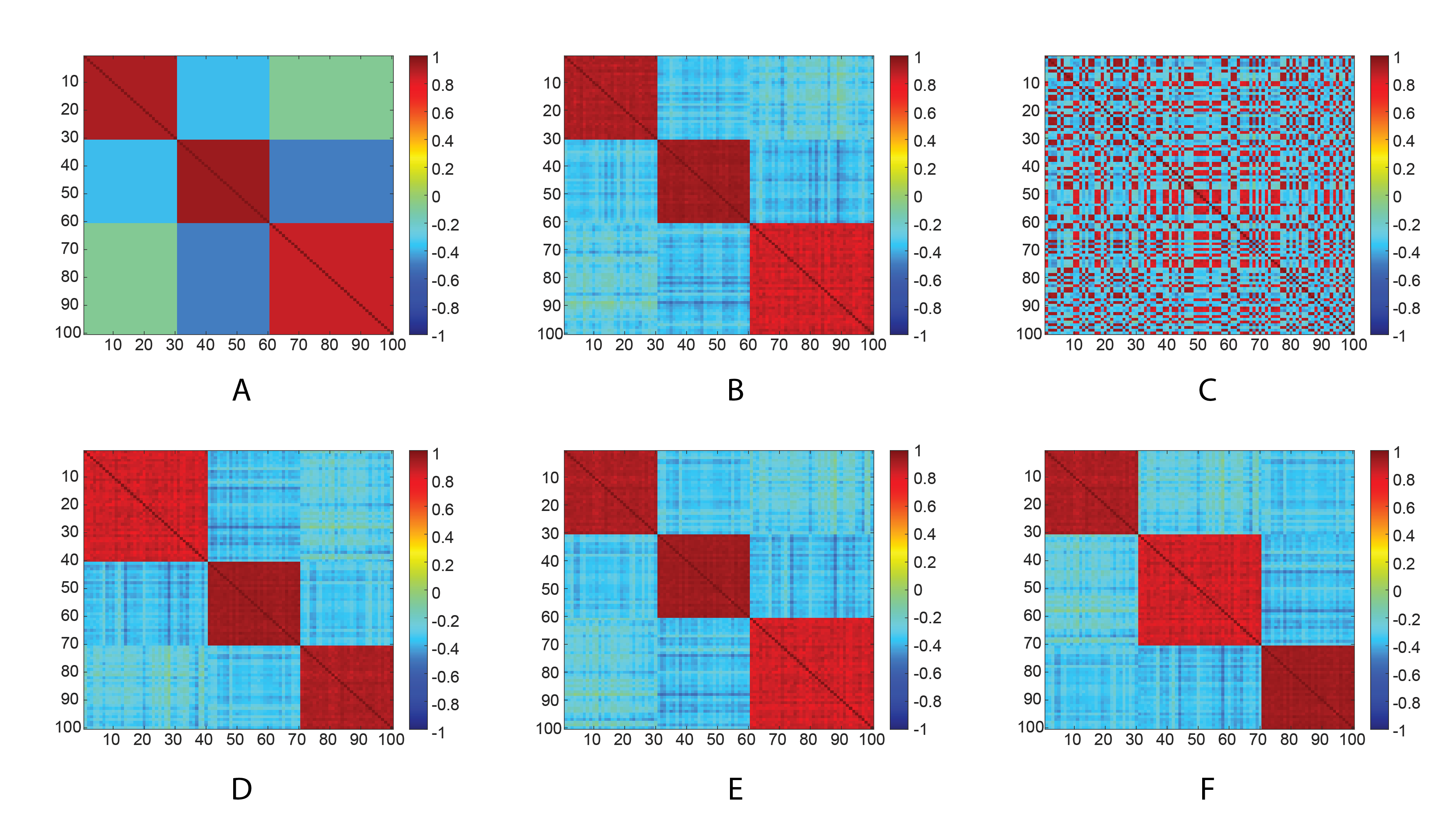} 
			\caption{
				We illustrate that various community detection algorithms extract almost identical patterns of the latent structure with $\bm{p}$.
				A: the heatmap of the population correlation matrix (``ground truth'') with $K = 3$ and community sizes of $p_1 = 30$, $p_2 = 30$, and $p_3 = 40$;
				B: we sample a multivariate dataset using the correlation matrix in A, resulting in the heatmap of a sample correlation matrix; 
				C: we permute the order of variables in the dataset, resulting in the heatmap of a sample correlation matrix with an unknown structure and unknown $\bm{p}$, which serves as the input for all community detection algorithms;
				D: the resulting heatmap to reveal the latent structure with $\bm{p}$ by applying \citet{WuMaLiu2021}'s algorithm to the sample correlation matrix in C;
				E: the resulting heatmap to reveal the latent structure with $\bm{p}$ by applying \citet{LiLeiBhattacharyya2022}'s algorithm to the sample correlation matrix in C;
				F: the resulting heatmap to reveal the latent structure with $\bm{p}$ by applying the $K$-medoids algorithm \citep{LeiRinaldo2015} to the sample correlation matrix in C.
			}
			\label{Fig:similar}
		\end{center}
	\end{figure}

	Recent developments of advanced network structure detection algorithms seem to provide an almost identical estimation of $\bm{p} = \left(p_1, \ldots, p_K\right) ^ \top$ when a latent organized pattern is present, as demonstrated in Figure~\ref{Fig:similar}.
	Therefore, it is plausible for the proposed approach to assume a known structure $\bm{p}$. 
	
	\section{PROOFS}
	
	\label{Sec:proof}
	
	\subsection{Proofs of Corollary B.1, Corollary B.2, and Corollary 1}
	
	\begin{proof}[Proof of Corollay B.1 and Corollay B.2]
		Please refer to the arguments and proofs in \citet{YangChenLee2024}.
	\end{proof}
	
	\begin{proof}[Proof of Corollary 1]
		
		Based on Definition~\ref{Dfn:ub}, a covariance matrix characterized by an interconnected community structure is a uniform-block matrix, therefore it can be expressed by $\bm{\Sigma} = \bm{\Sigma}\left(\textbf{A}, \textbf{B}, \bm{p} \right)$.
		Then, 
		\begin{equation*}
			\bm{\Sigma} \left(\textbf{A}, \textbf{B}, \bm{p} \right) 
			= \operatorname{Bdiag} \left(\bm{\ell}_1, \ldots, \bm{\ell}_K\right) \times \bm{\Sigma}_{\bm{f}} \times 
			\operatorname{Bdiag} \left(\bm{\ell}_1 ^ \top, \ldots, \bm{\ell}_K ^ \top \right) + \bm{\Sigma}_{\bm{u}}.
		\end{equation*}
		Using the block Hadamard product representation of a uniform-block matrix, we rewrite it:
		\begin{equation*}
			\begin{split}
				\bm{\Sigma}\left(\textbf{A}, \textbf{B}, \bm{p}\right)
				& = \textbf{A} \circ \textbf{I}(\bm{p}) + \textbf{B} \circ \textbf{J}(\bm{p})
				= \operatorname{Bdiag}\left(a_{11}\textbf{I}_{p_1}, \ldots, a_{K,K}\textbf{I}_{p_K}\right)
				+ \left(b_{kk'} \textbf{1}_{p_k \times p_{k'}}\right) \\
				& = \operatorname{Bdiag} \left(\bm{\ell}_1, \ldots, \bm{\ell}_K\right) \times \bm{\Sigma}_{\bm{f}} \times 
				\operatorname{Bdiag} \left(\bm{\ell}_1 ^ \top, \ldots, \bm{\ell}_K ^ \top \right) + \bm{\Sigma}_{\bm{u}},
			\end{split}
		\end{equation*}
		where matrices $\textbf{A} = \operatorname{diag}\left(a_{11}, \ldots, a_{KK}\right)$ and $\textbf{B} = \left(b_{kk'}\right)$ with $b_{k'k} = b_{kk'}$ for every $k \neq k'$ are known, 
		$\bm{\Sigma}_{\bm{f}} = (\sigma_{\bm{f}, kk'}) \in \mathbb{R} ^ {K \times K}$ is an unknown symmetric matrix with $\sigma_{\bm{f}, k'k} = \sigma_{\bm{f}, k'k}$ for every $k \neq k'$, 
		and $\bm{\Sigma}_{\bm{u}} = \operatorname{diag}\left(\sigma_{\bm{u}, 11}, \ldots, \sigma_{\bm{u}, pp} \right) \in \mathbb{R} ^ {p \times p}$ is an unknown diagonal matrix.
		
		Do some algebraic, we have that 
		\begin{equation*}
			\begin{split}
				& \operatorname{Bdiag} \left(\bm{\ell}_1, \ldots, \bm{\ell}_K\right) \times \bm{\Sigma}_{\bm{f}} \times 
				\operatorname{Bdiag} \left(\bm{\ell}_1 ^ \top, \ldots, \bm{\ell}_K ^ \top \right) + \bm{\Sigma}_{\bm{u}} \\
				& = \begin{pmatrix}
					\bm{\ell}_1 & \ldots & \textbf{0}_{p_1 \times 1} \\
					\textbf{0}_{p_2 \times 1} & \ldots & \textbf{0}_{p_2 \times 1} \\
					& \ddots & \\
					\textbf{0}_{p_K \times 1} & \ldots & \bm{\ell}_K
				\end{pmatrix}_{p \times K}
				\begin{pmatrix}
					\sigma_{\bm{f}, 11} & \ldots & \sigma_{\bm{f}, 1K} \\
					\sigma_{\bm{f}, 21} & \ldots & \sigma_{\bm{f}, 2K} \\
					\vdots & \ddots & \vdots \\
					\sigma_{\bm{f}, K1} & \ldots & \sigma_{\bm{f}, KK}
				\end{pmatrix}_{K \times K}
				\begin{pmatrix}
					\bm{\ell}_1 ^ \top & \ldots & \textbf{0}_{1 \times p_K} \\
					\textbf{0}_{1 \times p_1} & \ldots & \textbf{0}_{1 \times p_K} \\
					& \ddots & \\
					\textbf{0}_{1 \times p_1} & \ldots & \bm{\ell}_K ^ \top
				\end{pmatrix}_{K \times p} + \bm{\Sigma}_{\bm{u}} \\
				& = \left(\left( \sigma_{\bm{f}, kk'} \bm{\ell}_k \bm{\ell}_{k'} ^ \top \right)_{p_k \times p_{k'}} \right) 
				+ \operatorname{diag}\left(\sigma_{\bm{u}, 11}, \ldots, \sigma_{\bm{u}, pp} \right).
			\end{split}
		\end{equation*}
		In other words, $\operatorname{Bdiag} \left(\bm{\ell}_1, \ldots, \bm{\ell}_K\right) \times \bm{\Sigma}_{\bm{f}} \times 
		\operatorname{Bdiag} \left(\bm{\ell}_1 ^ \top, \ldots, \bm{\ell}_K ^ \top \right) + \bm{\Sigma}_{\bm{u}}$ has the diagonal blocks $\left( \sigma_{\bm{f}, kk} \bm{\ell}_k \bm{\ell}_{k} ^ \top \right)_{p_k \times p_{k}} + \operatorname{diag}\left(\sigma_{\bm{u}, \bar{p}_{k - 1} + 1, \bar{p}_{k - 1} + 1}, \ldots, \sigma_{\bm{u}, \bar{p}_k, \bar{p}_k} \right)$ for every $k$ and the off-diagonal blocks $\left( \sigma_{\bm{f}, kk'} \bm{\ell}_k \bm{\ell}_{k'} ^ \top \right)_{p_k \times p_{k'}}$ for every $k \neq k'$.
		
		Since $\bm{\Sigma}\left(\textbf{A}, \textbf{B}, \bm{p}\right)$ has the diagonal block $a_{kk}\textbf{I}_{p_k} + b_{kk}\textbf{J}_{p_k}$ for every $k$ and the off-diagonal blocks $b_{kk'} \textbf{1}_{p_k \times p_{k'}}$ for every $k \neq k'$, by the definition of the equality of two matrices, 
		we obtain that 
		\begin{equation*}
			\begin{split}
				a_{kk}\textbf{I}_{p_k} + b_{kk}\textbf{J}_{p_k} & = \sigma_{\bm{f}, kk} \bm{\ell}_k \bm{\ell}_{k} ^ \top  + \operatorname{diag}\left(\sigma_{\bm{u}, \bar{p}_{k - 1} + 1, \bar{p}_{k - 1} + 1}, \ldots, \sigma_{\bm{u}, \bar{p}_k, \bar{p}_k} \right), \quad k = k' = 1, \ldots, K \\
				b_{kk'} \textbf{1}_{p_k \times p_{k'}} & = \sigma_{\bm{f}, kk'} \bm{\ell}_k \bm{\ell}_{k'} ^ \top, 
				\quad k \neq k', \quad k, k' = 1, \ldots, K.
			\end{split}
		\end{equation*}

		First, we recall that the first element of $\bm{\ell}_k$, i.e., $\ell_{\bar{p}_{k - 1} + 1, k} = \tau_k \neq 0$, for every $k$.
		
		Second, for fixed $k$, the equalities of diagonal elements yield $a_{kk} + b_{kk} = \sigma_{\bm{u}, mm} + \sigma_{\bm{f}, kk} \ell_{m,k} ^ 2$ for all $m = \bar{p}_{k - 1} + 1, \ldots, \bar{p}_k$, 
		while the equalities of off-diagonal elements yield $b_{kk} = \sigma_{\bm{f}, kk} \ell_{m,k} \ell_{m',k}$ for all pairs $m \neq m' = \bar{p}_{k - 1} + 1, \ldots, \bar{p}_k$.
		
		(case 1) We focus on the off-diagonal equalities. 
		If $\sigma_{\bm{f}, kk} = 0$, then $b_{kk} = 0$ but this is impossible because $\textbf{B} \succ 0$,
		by Hadamard's inequality, $b_{kk} \neq 0$.
		If $\sigma_{\bm{f}, kk} \neq 0$, we define $b_{kk} ^ * = b_{kk} / \sigma_{\bm{f}, kk}$ thus $b_{kk} ^ * = \ell_{m,k} \ell_{m',k}$ for all pairs $m \neq m' = \bar{p}_{k - 1} + 1, \ldots, \bar{p}_k$.
		Taking $m = \bar{p}_{k - 1} + 1$ and $m' \neq m$, we obtain that $b_{kk} ^ * = \tau_k \ell_{m',k}$.
		Fixing $m'$, and due to $p_k > 2$, taking another $m'' \in \{\bar{p}_{k - 1} + 1, \ldots, \bar{p}_k\}$ while $m'' \neq m'$ and $m'' \neq m$, we obtain that $b_{kk} ^ * = \tau_k \ell_{m'',k}$.
		By $b_{kk} ^ * = \ell_{m', k} \ell_{m'',k} = (b_{k,k} ^ * / \tau_k) ^ 2$, 
		we derive $b_{kk} ^ * = \tau_k ^ 2$, or equivalently, $\ell_{m, k} = b_{kk} ^ * / \tau_k = \tau_k$ for all $m = \bar{p}_{k - 1} + 1, \ldots, \bar{p}_k$ and $\sigma_{\bm{f}, kk} = b_{kk} / \tau_k ^ 2$.
		
		(case 2) We focus on the diagonal equalities.
		Since $a_{kk} + b_{kk} = \sigma_{\bm{u}, mm} + \sigma_{\bm{f}, kk} \ell_{m,k} ^ 2 = \sigma_{\bm{u}, mm} + \sigma_{\bm{f}, kk} \tau_k ^ 2 = \sigma_{\bm{u}, mm} + b_{k,k}$,
		we have $\sigma_{\bm{u}, mm} = a_{k,k}$ for all $m = \bar{p}_{k - 1} + 1, \ldots, \bar{p}_k$.
		
		Third, for fixed pair $k \neq k'$, if $\sigma_{\bm{f}, kk'} = 0$, then $b_{kk'} \textbf{1}_{p_k \times p_{k'}} = \textbf{0}_{p_k \times p_{k'}}$, which implies that $b_{kk'} = 0 = \sigma_{\bm{f}, kk'}$.
		If $\sigma_{\bm{f}, kk'} \neq 0$, then we define $b_{kk'} ^ * = b_{kk'} / \sigma_{\bm{f}, kk'}$, therefore, 
		$b_{kk'} ^ * \textbf{1}_{p_k \times p_{k'}} = \bm{\ell}_k \bm{\ell}_{k'} ^ \top = \left(\ell_{m, k} \ell_{m', k'}\right)$, or equivalently, $b_{kk'} ^ * = \ell_{m, k} \ell_{m', k'}$ for all pairs $m = \bar{p}_{k - 1} + 1, \ldots, \bar{p}_k$ and $m' = \bar{p}_{k' - 1} + 1, \ldots, \bar{p}_{k'}$.
		By the result in (case 1), $b_{kk'} ^ * = \ell_{m, k} \ell_{m', k'} = \tau_k \tau_{k'}$ for all pairs $m$ and $m'$.
		So, we have $b_{kk'} ^ * = \tau_k \tau_{k'}$, or equivalently, $\sigma_{\bm{f}, kk'} =  b_{kk'} / (\tau_k \tau_{k'})$.
		
		In summary, we conclude that $\bm{\ell}_k = \tau_k \textbf{1}_{p_k \times 1}$ for every $k$, 	
		$\bm{\Sigma}_{\bm{f}} = \operatorname{diag}\left(\tau_1 ^ {-1}, \ldots, \tau_K ^ {-1}\right) \times \textbf{B} \times \operatorname{diag}\left(\tau_1 ^ {-1}, \ldots, \tau_K ^ {-1}\right)$,
		and $\bm{\Sigma}_{\bm{u}} = \textbf{A} \circ \textbf{I}(\bm{p})$.
	\end{proof}
	
	\subsection{Proofs of Theorem 1, Theorem 2, and Theorem 3}
	
	\begin{proof}[Proof of Theorem 1]
		
		We sketch the steps of proof here and please refer to more details in the arguments and proof for Theorem 1 in \citet{YangChenChen2024}.
		
		As mentioned in the manuscript, without loss of generality, we set $\tau_k = 1$ for each $k$. 
		Thus, using the results in Corollary 1, 
		the covariance matrix, which is a uniform-block matrix $\bm{\Sigma}\left(\textbf{A}, \textbf{B}, \bm{p}\right)$, can be decomposed into $\bm{\Sigma}\left(\textbf{A}, \textbf{B}, \bm{p}\right) = \textbf{L} \bm{\Sigma}_{\bm{f}} \textbf{L} ^ \top + \bm{\Sigma}_{\bm{u}}$.
		In other words, we can analytically solve out $\textbf{L}$, $\bm{\Sigma}_{\bm{f}}$, and $\bm{\Sigma}_{\bm{u}}$ in terms of components of $\textbf{A}$, $\textbf{B}$, and $\bm{p}$:		
		$\textbf{L} = \operatorname{Bdiag}\left(\textbf{1}_{p_1 \times 1}, \ldots, \textbf{1}_{p_K \times 1} \right)$, $\bm{\Sigma}_{\bm{f}} = \left(b_{kk'}\right) = \textbf{B}$, and $\bm{\Sigma}_{\bm{u}} = \operatorname{diag}\left(a_{11}\textbf{I}_{p_1}, \ldots, a_{KK} \textbf{I}_{p_K}\right) = \textbf{A} \circ \textbf{I}(\bm{p})$. 
		
		Step 1, let $\bm{\theta} = \left(a_{11}, \ldots, a_{KK}, b_{11}, b_{12}, \ldots, b_{KK}\right) ^ \top$ denote the $q = K + (K + 1)K / 2$ dimensional unknown parameter vector.
		Thus, following the arguments in \citet{YangChenChen2024}, we can prove the score equation has a unique root, which is the maximum likelihood estimator $\widehat{\bm{\theta}} = \left(\widehat{a}_{11}, \ldots, \widehat{a}_{KK}, \widehat{b}_{11}, \widehat{b}_{12}, \ldots, \widehat{b}_{KK}\right) ^ \top$, 
		where $\widehat{a}_{kk}$ and $\widehat{b}_{kk'}$ have the closed-form expressions in (4).
		
		Step 2, let $\left(\textbf{S}_{kk'}\right)$ denote the partitioned matrix of $\textbf{S}$ by $\bm{p}$.		
		We define $\widehat{\alpha}_{kk} = p_k ^ {-1} \operatorname{tr}\left(\textbf{S}_{kk}\right)$, $\widehat{\beta}_{kk} = p_k ^ {-2} \operatorname{sum}\left(\textbf{S}_{kk}\right)$, and $\widehat{\beta}_{kk'} = (p_k p_{k'}) ^ {- 1} \operatorname{sum}\left(\textbf{S}_{kk'}\right)$ for every $k$ and every $k' \neq k$.
		Following the similar arguments in \citet{YangChenChen2024}, they are UMVUEs of 
		$\alpha_{kk} = p_k ^ {-1} \operatorname{tr}\left(\bm{\Sigma}_{kk}\right)$, $\beta_{kk} = p_k ^ {-2} \operatorname{sum}\left(\bm{\Sigma}_{kk}\right)$, and $\beta_{kk'} = (p_k p_{k'}) ^ {- 1} \operatorname{sum}\left(\bm{\Sigma}_{kk'}\right)$, respectively.
		
		Step 3, there is a linear transformation between $\widehat{a}_{kk}$ and $\widehat{b}_{kk'}$ between $\widehat{\alpha}_{kk}$ and $\widehat{\beta}_{kk'}$. Thus, $\widehat{a}_{kk}$ and $\widehat{b}_{kk'}$ are UMVUEs of $a_{kk}$ and $b_{kk'}$ for every $k$ and every $k' \neq k$.
		Therefore, the plug-in estimators $\widehat{\bm{\Sigma}}_{\bm{f}} = \left(\widehat{b}_{kk'}\right) = \widehat{\textbf{B}}$, 
		$\widehat{\bm{\Sigma}}_{\bm{f}} = \widehat{\textbf{A}} \circ \textbf{I}(\bm{p})$, 
		and the analytical solution $\widehat{\textbf{L}} = \operatorname{Bdiag}\left(\textbf{1}_{p_1 \times 1}, \ldots, \textbf{1}_{p_K \times 1}\right)$ are UMVUEs, as long as $K + (K + 1)K / 2 < n$ and both $K$ and $n$ are fixed.
		
		In addition to the finite-sample properties of $\bm{\theta}$, the large-sample properties, including consistency, asymptotic efficiency, and asymptotic normality, can be established. 
	\end{proof}
	
	\begin{proof}[Proof of Theorem 2]
		
		We verify Condition (A) in Theorem 1 proposed by \citet{LuSchmidt2012} to show the equivalence between the OLS, the GLS, and the FGLS of factor score $\bm{f}_i$.
		
		We start with $\bm{X}_i = \widehat{\textbf{L}} \bm{f}_i + \bm{u}_i$, 
		where the $p$ by $K$ matrix $\widehat{\textbf{L}} = \operatorname{Bdiag}\left(\textbf{1}_{p_1 \times 1}, \ldots, \textbf{1}_{p_K \times 1} \right)$ and $\bm{u}_i \sim N \left(\textbf{0}_{p \times 1}, \bm{\Sigma}_{\bm{u}} \right)$ with $\bm{\Sigma}_{\bm{u}} = \operatorname{diag}\left(a_{11} \textbf{I}_{p_1}, \ldots, a_{KK} \textbf{I}_{p_K} \right)$.
		Let $\textbf{P} = \operatorname{diag}\left(p_1, \ldots, p_K \right)$.
		
		Theorem 1 proposed by \citet{LuSchmidt2012} requires:
		$\bm{\Sigma}_{\bm{u}} \succ 0$, $\widehat{\textbf{L}} ^ \top \widehat{\textbf{L}} = \textbf{P} \succ 0$, and $\widehat{\textbf{L}} ^ \top \bm{\Sigma}_{\bm{u}} ^ {-1} \widehat{\textbf{L}} = \operatorname{diag}\left(p_1 / a_{11}, \ldots, p_K / a_{KK}\right) = \textbf{A} ^ {- 1} \textbf{P} \succ 0$, 
		which are satisfied. 
		Moreover, 
		\begin{equation*}
			\begin{split}
				\left(\widehat{\textbf{L}} ^ \top \widehat{\textbf{L}}\right) ^ {- 1}
				\widehat{\textbf{L}} ^ \top \bm{\Sigma}_{\bm{u}} \widehat{\textbf{L}}
				\left(\widehat{\textbf{L}} ^ \top \widehat{\textbf{L}}\right) ^ {- 1}
				& = \textbf{P} ^ {- 1} (\textbf{A} \textbf{P}) \textbf{P} ^ {- 1} 
				= \textbf{A} \textbf{P} ^ {- 1}, \\
				\left(\widehat{\textbf{L}} ^ \top \bm{\Sigma}_{\bm{u}} ^ {-1} \widehat{\textbf{L}}\right) ^ {- 1}
				& = (\textbf{A} ^ {- 1} \textbf{P}) ^ {- 1}
				= \textbf{P} ^ {- 1} \textbf{A}.
			\end{split}
		\end{equation*}
		Since both $\textbf{A}$ and $\textbf{P}$ are diagonal, $\left(\widehat{\textbf{L}} ^ \top \widehat{\textbf{L}}\right) ^ {- 1}
		\widehat{\textbf{L}} ^ \top \bm{\Sigma}_{\bm{u}} \widehat{\textbf{L}}
		\left(\widehat{\textbf{L}} ^ \top \widehat{\textbf{L}}\right) ^ {- 1} = \left(\widehat{\textbf{L}} ^ \top \bm{\Sigma}_{\bm{u}} ^ {-1} \widehat{\textbf{L}}\right) ^ {- 1}$,
		Condition (A) in Theorem 1 proposed by \citet{LuSchmidt2012} is satisfied. 		
		Therefore, Condition (C) in Theorem 1 holds. 
		In addition to $\bm{\Sigma_{\bm{u}}}$,
		we can check Condition (C) in Theorem 1 holds for $\widehat{\bm{\Sigma}}_{\bm{u}}$.
		So, we complete the equivalence between the OLS, the GLS, and the FGLS estimators of factor score $\bm{f}_i$.
		
		Additionally, $\widehat{\bm{f}}_i$ is UMVUE, normal distributed, and exhibits large-sample properties because it is the OLS estimator of a classical linear regression model with normal errors. 
	\end{proof}
	
	\begin{proof}[Proof of Theorem 3]
		
		To establish (1) in Theorem 3, i.e., the exact variance estimators of $\widehat{a}_{kk}$ and $\widehat{b}_{kk'}$ for every $k$ and every $k' \neq k$, 
		please refer to the arguments and proof for Corollary A.2 in the supplementary material proposed by \citet{YangChenChen2024}.
		
		To show (2), we compute the exact covariance matrix of $\widehat{f}_i$ as below:
		\begin{equation*}
			\begin{split}
				\operatorname{cov}\left(\widehat{\bm{f}}_i\right)
				& = \left(\widehat{\textbf{L}} ^ \top \widehat{\textbf{L}}\right) ^ {-1}
				\widehat{\textbf{L}} ^ \top
				\operatorname{cov} \left(\bm{X}_i \right) \left[\left(\widehat{\textbf{L}} ^ \top \widehat{\textbf{L}}\right) ^ {-1}
				\widehat{\textbf{L}} ^ \top\right] ^ \top \\
				& = \left(\widehat{\textbf{L}} ^ \top \widehat{\textbf{L}}\right) ^ {-1}
				\widehat{\textbf{L}} ^ \top
				\left(\textbf{L} \bm{\Sigma}_{\bm{f}} \textbf{L} ^ \top + \bm{\Sigma}_{\bm{u}} \right) \left[\widehat{\textbf{L}} \left(\widehat{\textbf{L}} ^ \top \widehat{\textbf{L}}\right) ^ {-1} \right] \\
				& = 
				\left(\widehat{\textbf{L}} ^ \top \widehat{\textbf{L}}\right) ^ {-1}
				\widehat{\textbf{L}} ^ \top \textbf{L} \bm{\Sigma}_{\bm{f}} \textbf{L} ^ \top \widehat{\textbf{L}} \left(\widehat{\textbf{L}} ^ \top \widehat{\textbf{L}}\right) ^ {-1} +
				\left(\widehat{\textbf{L}} ^ \top \widehat{\textbf{L}}\right) ^ {-1}
				\widehat{\textbf{L}} ^ \top \bm{\Sigma}_{\bm{u}} \widehat{\textbf{L}} \left(\widehat{\textbf{L}} ^ \top \widehat{\textbf{L}}\right) ^ {-1} \\
				& = \bm{\Sigma}_{\bm{f}}
				+ \textbf{P} ^ {-1}
				\textbf{A} \textbf{P} \textbf{P} ^ {-1} \\
				& = \textbf{B} + \textbf{P} ^ {-1}
				\textbf{A} \\
				& = \left(b_{kk'}\right) + \operatorname{diag}\left(
				a_{11} / p_{1}, \ldots, a_{KK} / p_K
				\right),
			\end{split}
		\end{equation*}
		where $\textbf{L} = \widehat{\textbf{L}} = \operatorname{Bdiag}\left(\textbf{1}_{p_1 \times 1}, \ldots, \textbf{1}_{p_K \times 1}\right)$ and $\bm{\Sigma}_{\bm{f}} = \textbf{B} = \left(b_{kk'}\right)$ with $b_{k'k} = b_{kk'}$.
	\end{proof}
	
	
	\bibliographystyle{imsart-nameyear} 
	\bibliography{paper_ref_SCFA}       
	